\documentclass{aa}  
\usepackage{graphicx}
\usepackage{txfonts}
\usepackage[colorlinks=true,citecolor=blue]{hyperref}
\usepackage{multirow}
\usepackage{amsmath,amstext}
\usepackage[T1]{fontenc}
\usepackage{color}
\usepackage{comment}
\usepackage{caption}
\DeclareCaptionFormat{cont}{#1 (cont.)#2#3\par}
\usepackage[flushleft]{threeparttable}

\DeclareRobustCommand{\ion}[2]{%
\relax\ifmmode
\ifx\testbx\f@series
{\mathbf{#1\,\mathsc{#2}}}\else
{\mathrm{#1\,\mathsc{#2}}}\fi
\else\textup{#1\,{\mdseries\textsc{#2}}}%
\fi}

\newcommand{\orcid}[1]{\href{https://orcid.org/#1}{\includegraphics[width=10pt]{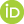}}}
\newcommand{\github}[1]{\href{https://github.com/#1}{\includegraphics[width=10pt]{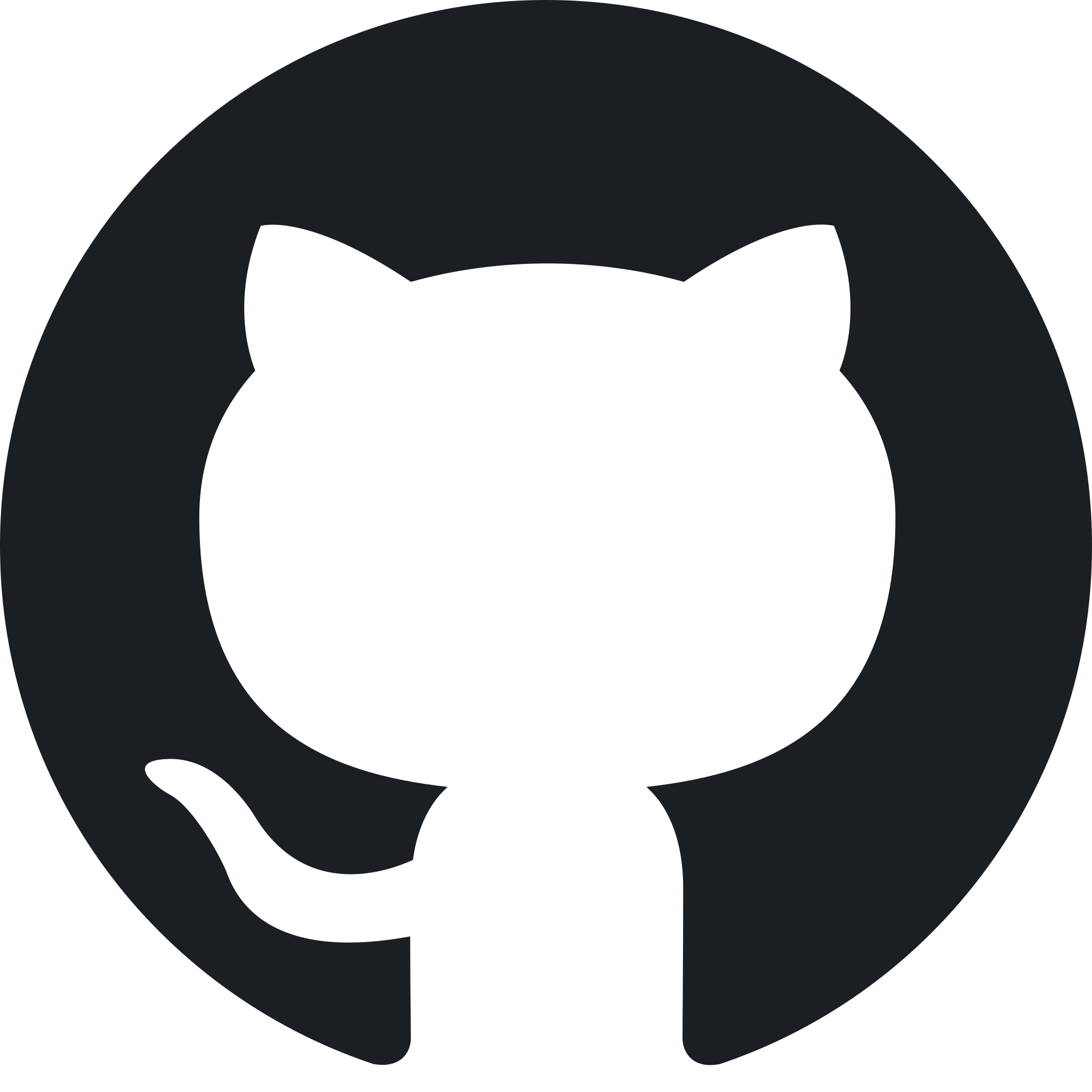}}}

\newcommand{\snia}{SN\,Ia\xspace}
\newcommand{\sneia}{SNe\,Ia\xspace}

\newcommand{\cspi}{CSP-I\xspace}
\newcommand{\cspii}{CSP-II\xspace}
\newcommand{\snoopy}{SNooPy\xspace}

\newcommand{\maxmodel}{\textit{max\_model}\xspace}
\newcommand{\EBVmodel}{\textit{EBV\_model2}\xspace}

\newcommand{\tmax}{$t_{\rm max}$\xspace}
\newcommand{\sbv}{$s_{BV}$\xspace}

\newcommand{\J}{\textit{J}\xspace}
\renewcommand{\H}{\textit{H}\xspace} 
\newcommand{\Y}{\textit{Y}\xspace} 
\newcommand{\nickel}{$^{56}$Ni\xspace}
\newcommand{\mni}{$M_{\rm Ni}$\xspace}
\newcommand{\mfe}{$M_{\rm Fe}$\xspace}
\newcommand{\ebvhost}{$E(B-V)_{\rm host}$\xspace}
\newcommand{\hostmass}{log$_{10}(M/M_{\odot})$\xspace}

\defcitealias{Kasen2006}{K06}

\newcommand{\hunits}{km\,s$^{-1}$\,Mpc$^{-1}$\xspace}

\begin{document} 

\title{Analyzing Type Ia supernovae near-infrared light curves with principal component analysis}
\titlerunning{\sneia NIR Analysis with PCA}
\authorrunning{M\"uller-Bravo et al.}

\author{
    T.~E.~M\"uller-Bravo\inst{\ref{ICE}, \ref{IEEC}, \ref{TCD}, \ref{ICEN}}\thanks{\email{t.e.muller-bravo@tcd.ie}}\orcid{0000-0003-3939-7167},
    L.~Galbany\inst{\ref{ICE}, \ref{IEEC}}\orcid{0000-0002-1296-6887},
    M. D. Stritzinger\inst{\ref{aarhus}}\orcid{0000-0002-5571-1833},
    C.~Ashall\inst{\ref{IfA}}\orcid{0000-0002-5221-7557},
    E.~Baron\inst{\ref{tucson},\ref{inst:hsde}}\orcid{0000-0001-5393-1608},
    C.~R.~Burns\inst{\ref{Carnegie}}\orcid{0000-0003-4625-6629},
    P.~Höflich\inst{\ref{Florida}},
    N.~Morrell\inst{\ref{LasCampanas}}\orcid{0000-0003-2535-3091},
    M.~Phillips\inst{\ref{LasCampanas}}\orcid{0000-0003-2734-0796},
    N.~B.~Suntzeff\inst{\ref{Texas}}\orcid{0000-0002-8102-181X},
    S.~A.~Uddin\inst{\ref{CASSA}, \ref{APUS}}\orcid{0000-0002-9413-4186}
}
\institute{
    Institute of Space Sciences (ICE, CSIC), Campus UAB, Carrer de Can Magrans, s/n, E-08193 Barcelona, Spain
    \label{ICE}
    \and Institut d’Estudis Espacials de Catalunya (IEEC), E-08034 Barcelona, Spain 
    \label{IEEC}
    \and School of Physics, Trinity College Dublin, The University of Dublin, Dublin 2, Ireland
    \label{TCD}
    \and Instituto de Ciencias Exactas y Naturales (ICEN), Universidad Arturo Prat, Chile
    \label{ICEN}
    \and Department of Physics and Astronomy, Aarhus University, Ny Munkegade 120, DK-8000 Aarhus C, Denmark
    \label{aarhus}
    \and Institute for Astronomy, University of Hawai'i, 2680 Woodlawn Drive, Honolulu HI 96822, USA
    \label{IfA}
    \and Planetary Science Institute, 1700 E Fort Lowell Rd., Ste 106, Tucson, AZ 85719 USA
    \label{tucson}
    \and Hamburger Sternwarte, Gojensbergweg 112, 21029 Hamburg, Germany
    \label{inst:hsde}
    \and Observatories of the Carnegie Institution for Science, 813 Santa Barbara Street, Pasadena, CA 91101, USA
    \label{Carnegie}
    \and Department of Physics, Florida State University, Tallahassee, 32306, USA
    \label{Florida}
    \and
    Las Campanas Observatory, Carnegie Observatories, Casilla 601, La Serena, Chile
    \label{LasCampanas}
    \and 
    George P. and Cynthia Woods Mitchell Institute for Fundamental Physics and Astronomy, Department of Physics and Astronomy, Texas A\&M University, College Station, TX 77843, USA 
    \label{Texas}
     \and 
     Center for Astronomy, Space Science and Astrophysics, Independent University, Bangladesh, Dhaka 1245, Bangladesh
     \label{CASSA}
    \and
     American Public University System, Charles Town, WV 25414, USA
    \label{APUS}  
}

\date{Received \today; accepted ---}

\abstract
{ Thermonuclear explosions of C/O white dwarf
stars in binary systems known as Type Ia supernovae (\sneia) remain poorly understood. The complexity of their progenitor systems, explosion physics, and intrinsic diversity poses challenges in understanding these phenomena as astrophysical objects, as well as their standardization and use as cosmological probes. 
Near-infrared (NIR) observations offer a promising avenue for studying the physics of \sneia and for reducing systematic uncertainties in distance estimations, as they exhibit lower dust extinction and smaller dispersion in peak luminosity than optical bands.
In this work, we applied a principal component analysis (PCA) to a sample of \sneia with well-sampled NIR (\textit{YJH}-band) light curves to identify the dominant components of their variability and constrain physical underlying properties.
The theoretical models are used for the physical interpretation of the PCA components, where we found that the \nickel mass best describes the dominant variability. Other factors, such as mixing and metallicity, were found to contribute significantly as well. However, some differences are seen among the components of the NIR bands, which could be attributed to differences in the explosion aspects they each trace.
Additionally, we compared the PCA components to various light curve parameters, identifying strong correlations between the first component in \J and \H bands (second component in \Y) and peak brightness in both the NIR and optical bands, particularly in the \Y band.
When applying a PCA to NIR color curves, we found interesting correlations with the host-galaxy mass, where \sneia with redder NIR colors are predominantly found in less massive (potentially more metal-poor) galaxies.
We also investigated the potential for improved standardization in the $Y$ band by incorporating PCA coefficients as correction parameters, leading to a reduction in the scatter of the intrinsic luminosity of \sneia. 
As new NIR observations become available, our findings can be further tested, ultimately refining our understanding of \sneia physics and enhancing their reliability as cosmological distance indicators.
}

\maketitle

\section{Introduction}

It has long been postulated that Type~Ia supernovae (\sneia) result from the thermonuclear explosions of carbon and oxygen (C/O) in white dwarfs (WDs) of binary systems \citep[e.g., ][] {Whelan1973, Iben1984, Woosley1986}. Although some constraints have been determined based on observations of nearby \sneia \citep[][]{Nugent2011}, conclusive evidence on the progenitor systems of these transients remains elusive.
The bulk population (i.e., omitting sub-types) of \sneia are well-known for being relatively homogeneous in luminosity. In the optical, brighter \sneia tend to have wider light curves \citep{Pskovskii1977, Phillips1993} and bluer colors \citep{Tripp1998}, while the case is vice-versa for dimmer SNeIa. These correlations provide useful insights into some physical properties of these objects. For instance, brighter \sneia also produce more $^{56}$Ni mass (\mni), which can be related to higher temperature and, therefore, bluer colors, on longer diffusion timescales and also feature longer-lasting luminosity emission \citep[e.g.,][]{Nugent1995}. 

Although, the number of \sneia observed in the optical has drastically increased in recent years, thanks to such surveys as Zwicky Transient Facility \citep{Bellm2019, Masci2019, Graham2019, Dekany2020} and in the near future with the Vera C. Rubin Observatory Legacy Survey of Space and Time \citep{Ivezic2019}, these wavelengths only provide a limited amount of information.
At near-infrared (NIR) wavelengths, \sneia are considered standard candles \citep[NIR; e.g.,][]{Elias1981, Elias1985, Meikle2000, Krisciunas2004, Avelino2019}. Their high degree of intrinsic homogeneity at these wavelengths even allows us to estimate their peak brightness with a single photometric epoch, with the requirement of well-sampled optical light curves \citep{Stanishev2018, Müller-Bravo2022}. However, this low dispersion only applies to their first peak. \sneia develop a distinct secondary NIR peak a few weeks after the first one, but the characteristics of it largely vary from object to object. Therefore, the study of \sneia at NIR wavelengths gives provides further insights into their progenitors and explosion mechanisms \citep[e.g.,][]{Wheeler1998, Kasen2006, Hsiao2013, Hsiao2015, Hsiao2019, Hoogendam2025}.

Some early theoretical modeling of \sneia at NIR wavelengths, such as that of  \cite{Hoeflich1995} and \cite{Pinto2000}, hinted toward a drop in opacity, that is, a decrease in the diffusion time, which would explain the secondary peak. This would also help us avoid using an additional power source to inject energy at later epochs. \cite{Pinto2000} also noted an increase in the abundance of \ion{Fe}{ii} coinciding with the increase in emissivity in the NIR. \citet[][hereafter K06]{Kasen2006}, also used theoretical modeling, coming to the conclusion that the secondary peak is produced by the recombination of iron-group elements inside the ejecta of the SN, going from doubly to singly ionised (e.g., \ion{Fe}{iii} $\rightarrow$ \ion{Fe}{ii}). The prominence and time of the peak was found to depend on several physical parameters, such as \mni, the metallicity of the progenitor, and the amount of mixing of \nickel.
In addition, \cite{Hoeflich2017} argued that the NIR flux reaches a secondary maximum when the radius of the photosphere reaches its maximum, assuming the NIR flux follows Wein's limit\footnote{For an expanding ejecta, the flux is proportional to $TR_{\rm phot}$, where $T$ is the temperature and $R_{\rm phot}$ is the radius of the photosphere.}.

Purely based on observations, \cite{Phillips2012} showed that brighter \sneia tend to have brighter and delayed secondary peaks, while fainter ones have fainter and earlier secondary peaks; even turning into a \lq shoulder' instead of a distinct secondary peak for the SNe in the faint end of the luminosity distribution (see also \citealt{Ashall2020} and \citealt{Pessi2022}, although for the \textit{i} band). \cite{Ashall2020} also showed that \sneia \textit{i}-band light curves present different behaviors for different \snia sub-types, most likely tracing different physical aspects of these transients. Additionally, studying how the environment of \sneia affects their light curves across multiple wavelengths also provides useful information about the physics of these transients \citep[e.g.,][]{Uddin2020, Johansson2021, Ponder2021, Grayling2024}.

In this work, we performed a qualitative study of \sneia to understand the characteristics of their NIR light curves. We made use of machine learning methods widely used by the astronomical community to study \sneia and link observational properties to their physical parameters.
 
The outline of this paper is as follows. Sections~\ref{sec:samples} \& \ref{sec:selection} describe the sample of \sneia used and the initial selection cuts. The light-curve fits and calculation of the NIR rest-frame light curves are described in Sect.~\ref{sec:rest_lcs}, followed by the decomposition method in Sect.~\ref{sec:lc_decomposition}. The analysis of the decomposition and correlation between different light-curve and physical parameters is presented in Sect.~\ref{sec:analysis},  with the summary and conclusions given in Sect.~\ref{sec:conclusions}.

\section{Type Ia supernova near-infrared samples}
\label{sec:samples}

Historically, \snia surveys have focused on optical observations, with only a few extending to the NIR, mainly due to a lack of detectors at these wavelengths. However, in recent years, as NIR array technology has significantly improved, and thanks to the effort of several surveys, the number of \sneia observed at NIR wavelengths has rapidly increased. The surveys included as part of this work are described below.\\

\noindent CSP: The Carnegie Supernova Project
\citep[CSP;][]{Hamuy2006} provides optical and NIR (\textit{uBgVriYJH} bands) observations of 349 SNe~Ia, split into two samples. \cspi contains 134 SNe~Ia from three public data releases \cite[][]{Contreras2010, Stritzinger2011, Krisciunas2017}. \cspii contains 215 SNe~Ia, described in \cite{Phillips2019} and \cite{Hsiao2019}, with an upcoming data release (Suntzeff et al. in prep.).\\

\noindent CfAIR2: The Harvard-Smithsonian Center for Astrophysics IR2 sample (CfAIR2) consists of several data releases with optical photometry \citep[CfA1-5][]{Riess1999, Jha2006, Hicken2009a, Hicken2012} and two with NIR photometry \citep[CfAIR1-2;][]{Wood-Vasey2008, Friedman2015}. This sample includes 94 SNe~Ia with \textit{UBVRIr'i'JH} photometry.\\

\noindent RATIR: The Reionization and Transients InfraRed sample \citep[RATIR; ][]{Johansson2021} presents optical and NIR observations (\textit{UBgVrRiIzZYJH} bands) of 42 SNe observed as part of the intermediate Palomar Transient Factory (iPTF) survey.\\

\noindent DEHVILS: The Dark Energy, $H_0$, and peculiar Velocities using Infrared Light from Supernovae \citep[DEHVILS;][]{Peterson2023} survey observed 96 SNe with NIR ($YJH$ bands) photometry. Optical photometry of these SNe was covered by the $c$ and $o$ bands of the Asteroid Terrestrial-Impact Last Alert System \citep[ATLAS][]{Tonry2018}. This is additionally complemented with ZTF $g$- and $r$-bands light curves for 78 of the 96 SNe using the Automatic Learning for the Rapid Classification of Events \citep[ALeRCE;][]{Förster2021} broker.\\

\noindent Literature: We conducted a literature search for \sneia with available rest-frame NIR photometry. The following objects are those that passed the initial selection cuts discussed in Sect.~\ref{sec:selection}: 
SNe~1998bu \citep{Meikle2000}, 
1999ac \citep{Phillips2006}, 
1999cl \citep{Krisciunas2006},
1999ee \citep{Krisciunas2004a}, 
2000E \citep{Valentini2003}, 
2001ba \citep{Krisciunas2004a}, 
2001bt \citep{Krisciunas2004a}, 
2001el \citep{Krisciunas2003}, 
2002bo \citep{Krisciunas2004b, Miknaitis2007}, 
2002cv \citep{Elias-Rosa2008}, 
2002dj \citep{Pignata2008, Miknaitis2007}, 
2002fk \citep{Cartier2014}, 
2003cg \citep{Elias-Rosa2006, Miknaitis2007}, 
2003du \citep{Stanishev2007},
2005df \citep{Krisciunas2017b}, 
SDSS3241, SDSS3331 \citep{Freedman2009},
2011fe \citep{Matheson2012},
2013dy \citep{Pan2015}, 
2014J \citep{Foley2014, Marion2015, Srivastav2016},
and 
2017cbv \citep{Wee2018, Wang2020}. 
We note that the photometry of SN~2013dy was published in the AB magnitude system, so we converted it to the Vega magnitude system following \cite{Galbany2023}\footnote{AB to Vega offsets of $0.894$ and $1.368$\,mag in \textit{J} and \textit{H} bands are applied, respectively. We did not find any published AB to Vega offset value for the \textit{Y} band, so we follow the notes in \url{https://ratir.astroscu.unam.mx/instrument.html} and apply an offset of $0.634$\,mag as described in Table~7 of \cite{Hewett2006}, which is only an approximation.}.

\section{Initial selection of supernovae}
\label{sec:selection}

Despite the relatively large number of \sneia with NIR data described in Sect.~\ref{sec:samples}, many objects have a limited number of photometric epochs. This study requires good data coverage of the light curves,  sampling the first and second peaks, as well as the local minimum between them (we also refer to this minimum as the "valley").

The first cut is meant to select spectroscopically classified SNe, followed by selecting those with NIR observations on any band. In addition, as several \sneia have photometry from both CSP and CfAIR2 (considered " duplicates"), we used optical photometry from CSP and combined the NIR photometry of both surveys to obtain better coverage. We note that the NIR photometry of these surveys is in very good agreement, as shown in \cite{Friedman2015}. The same approach was used for duplicated SNe between CSP and RATIR, favoring CSP optical photometry. We then required the \sneia to have optical coverage to perform light-curve fits (see further below) and, finally, we only considered those classified as normal.

To get a relatively accurate estimation of the NIR light-curve shape, particularly around the secondary peak, a minimum of five epochs are needed: ideally before the first peak, around the first peak, around the valley, around the second peak, and after the second peak.
We then fit the \sneia light curves with \snoopy \citep{Burns2011, Burns2014} using the \maxmodel model (note: these fits are also used in the analysis in Sect.~\ref{sec:analysis}). 
The last criterium used is to select all the \sneia that have at least one NIR epoch before optical peak and one epoch 20 days after optical peak on any NIR band, independently.

\begin{table*}[!h]
\caption{First set of cuts applied to the sample of SNe~Ia (Section~\ref{sec:selection}).}
\centering

\setlength{\tabcolsep}{10pt}

\begin{tabular}{lccccccccc}
\hline
Cut & CSP & CfAIR2  & RATIR & DEHVILS & Literature & Total \\
\hline
Before cuts & 349 & 94 & 42 & 96 & 21 & 602 \\
Classified & 349 & 94 & 41 & 84 & 21 & 589 \\
NIR         & 302 & 94 & 40 & 84 & 21 & 541 \\
Duplicates & 302 & 69 & 30 & 84 & 21 & 506 \\
Optical     & 302 & 55 & 30 & 84 & 21 & 492 \\
Normal Ia & 245 & 52 & 25 & 83 & 21 & 426 \\
$\geq5$ NIR epochs & 173 & 49 & 13 & 75 & 21 & 331 \\
NIR LC coverage & 61 & 19 & 7 & 50 & 21 & 158 \\
\hline
\end{tabular}

\begin{tablenotes}
 \item \textbf{Notes.} By construction, the literature sample already includes all  cuts applied to the other samples.\\ 
\end{tablenotes}
\label{tab:cuts}

\end{table*}

These minimal selection cuts, presented in Table~\ref{tab:cuts}, reduce the initial sample to 158 normal \sneia with five NIR epochs on any band and with a partial coverage of the first and second peaks. We note that this number is further reduced in the following sections due to other selection criteria.

\section{Rest-frame light curves}
\label{sec:rest_lcs}

\begin{figure}[ht]
    \includegraphics[width=\columnwidth]{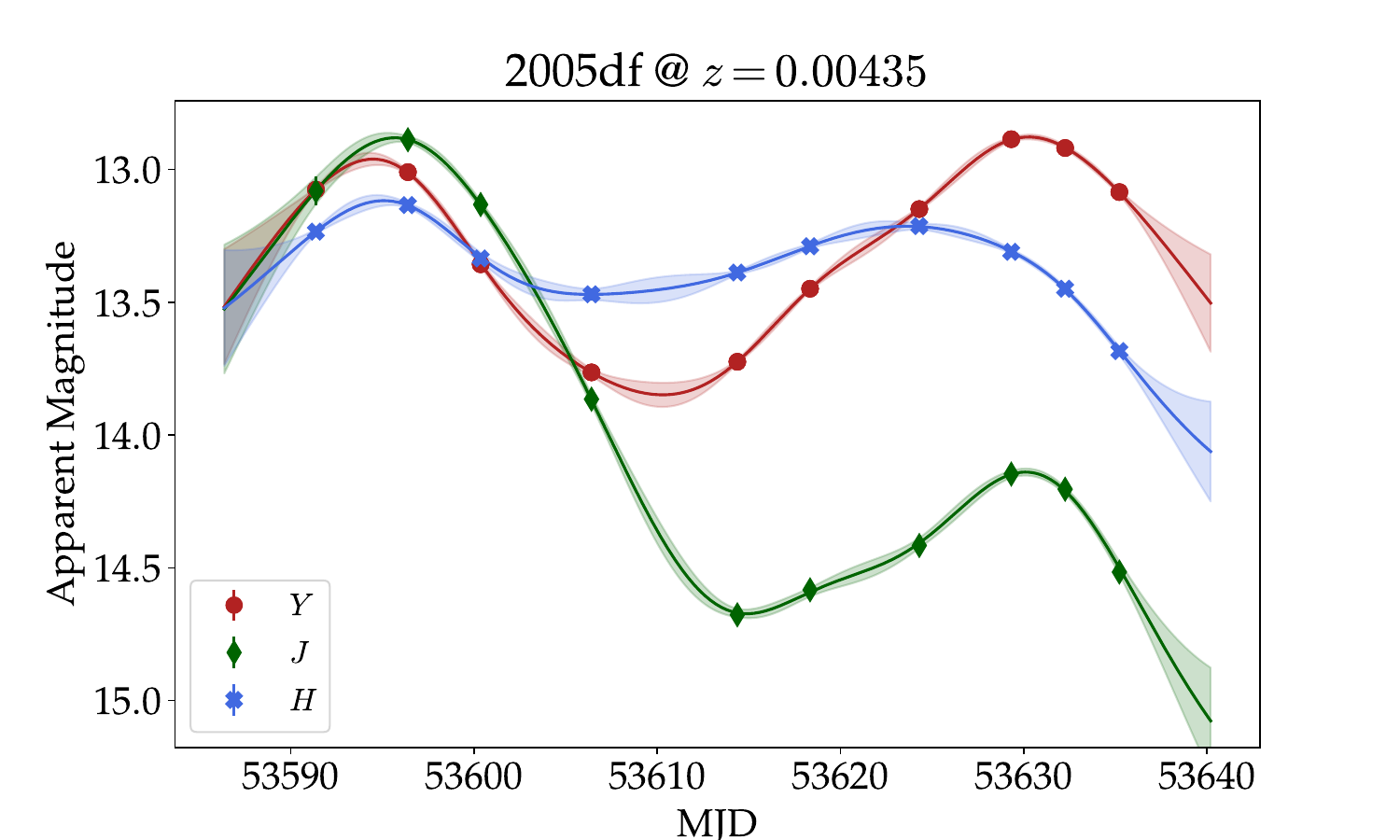}
    \includegraphics[width=\columnwidth]{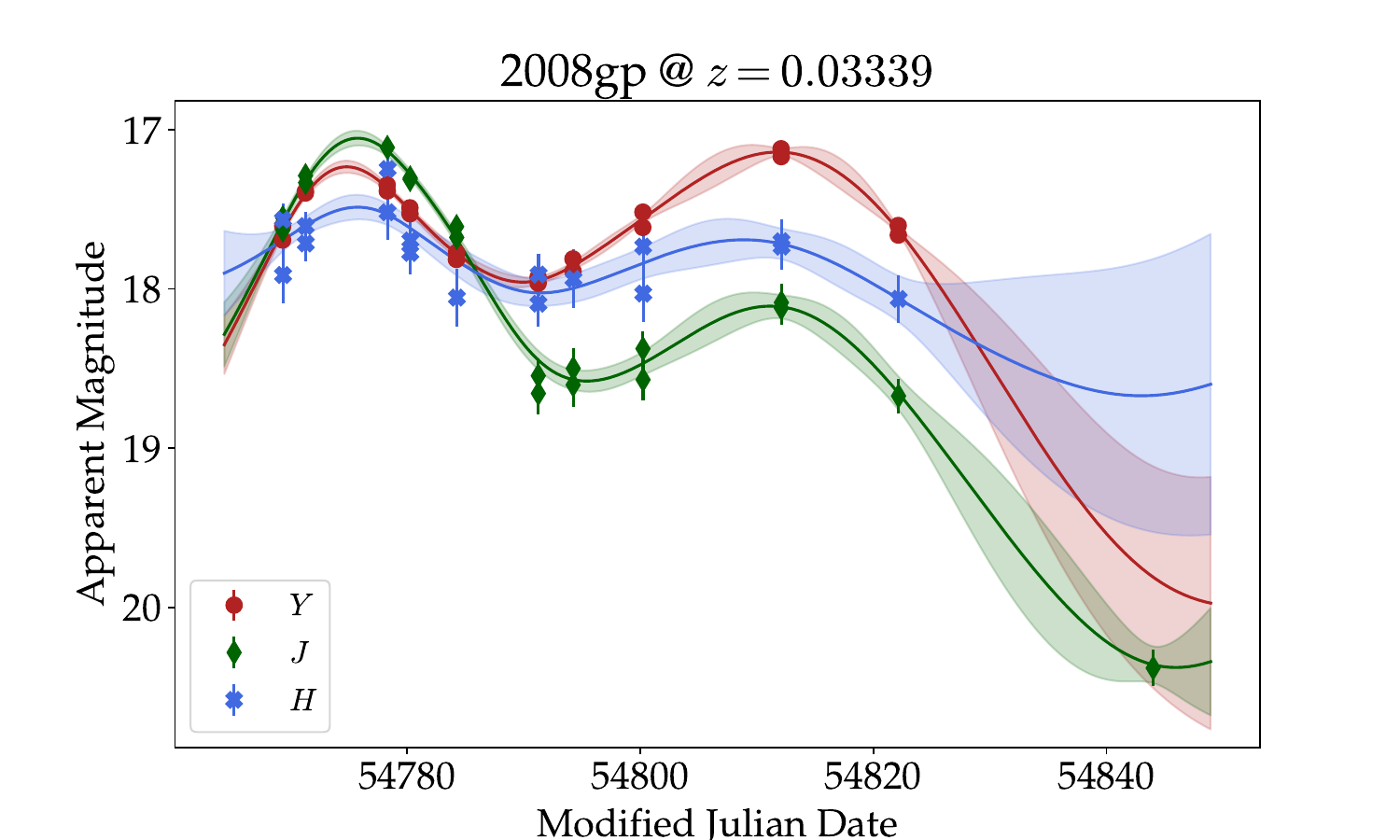}
    \includegraphics[width=\columnwidth]{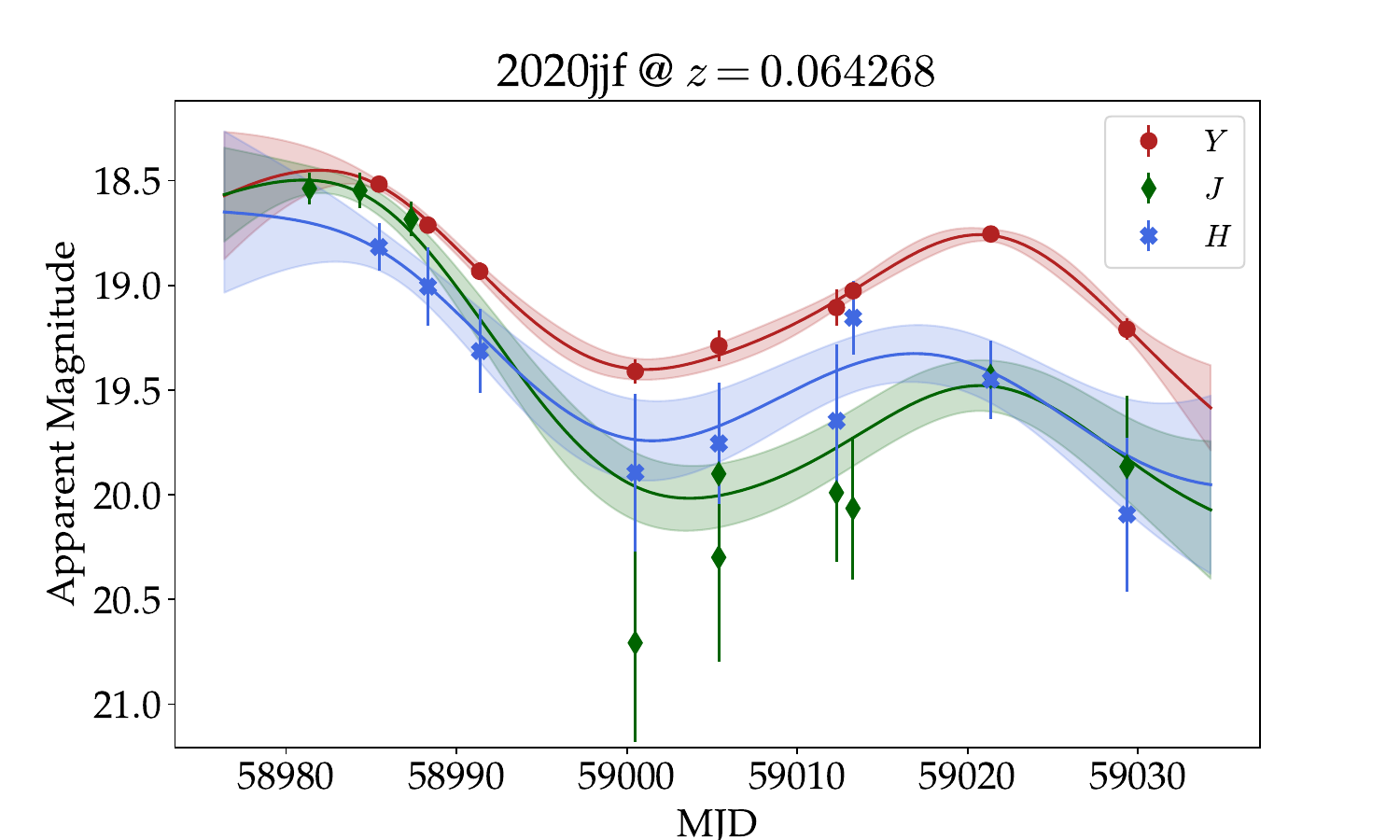}
    \caption{PISCOLA light-curve (\textit{YJH}-bands) fits of three SNe~Ia: 2005df (Literature; top), 2008gp (CSP; middle), and 2020jjf (DEHVILS; bottom). The photometric uncertainties propagates into the fits themselves.
    }
    \label{fig:lcs_fits}
\end{figure}

To obtain the rest-frame light curves, $K$-corrections \citep{Oke1968} are commonly used. This normally requires knowing the spectral energy distribution (SED) of the object in question, for which previously trained templates can be used \citep[e.g.,][]{Hsiao2007}. 
In this work, we follow a similar approach as in \cite{Pessi2022} and \cite{Galbany2023}. The observed photometry was first fit with \snoopy using the \maxmodel model (as mentioned in Sect.~\ref{sec:selection}), which outputs \textit{K}-corrected and Milky-Way dust-extinction corrected photometry. We used version \textsc{2.7.0} of \snoopy that includes the updated NIR SED templates from \cite{Lu2023}. Then, the corrected photometry was fit with PISCOLA \citep[version \textsc{3.0.0};][]{piscola}, a data-driven light-curve fitter based on Gaussian process regression \citep[GPR;][]{Rasmussen2006}, to produce continuous, rest-frame light curves. As PISCOLA fits multiple bands at the same time, information across light curves is used to interpolate the gaps in the data. 
Examples of PISCOLA light-curve fits to the corrected photometry are shown in Fig.~\ref{fig:lcs_fits}. Each \snia has a different cadence and signal-to-noise ratio (S/N), which is clearly propagated into the fit's uncertainty. To summarize, \snoopy is used to apply accurate \textit{K}-corrections at the epochs of observations, while PISCOLA provides a smooth interpolation and extrapolation of these.

\section{Near-infrared light-curve decomposition}
\label{sec:lc_decomposition}

Principal component analysis \citep[PCA; ][]{Pearson1901, Hotelling1933} is a statistical technique for linear transformation of a set of basis functions into a new coordinate system. The new basis functions are sorted in terms of the degree to which they describe the variance in the original data. Although PCA is commonly used for dimensionality reduction, it has several different applications depending on the goal. In the field of supernovae, PCA has been used in many different studies \citep[e.g.,][]{Cormier11, Galbany2014, Ishida2014, He18, Shahbandeh2022, Holmbo2023, Lu2023, Burrow2024}. In this work, PCA is used to produce a qualitative analysis of \sneia NIR light curves.

\subsection{Light-curve preprocessing}
\label{subsec:preprocessing}

Before applying PCA, a series of steps need to be performed for producing a sensible analysis. First, as PISCOLA fits do not rely on templates, we visually inspected the rest-frame light curves to make sure they are well-sampled and have physically expected shapes (i.e.,  two distinct peaks and a relatively smooth shape). This step mainly allows us to avoid including any noise\, in the decomposition. Primarily, SNe were removed due to relatively poor fit quality (e.g. due to large data uncertainty) or partial coverage of the light curves (2006D,
2008hy,
2008O,
2020kqv,
2020krw,
2020mby,
2020mdd,
2020tfc,
and 2021glz).
Additionally, SNe 2006le and 2009Y were removed only from the \H-band light-curve analysis due to poor fit quality as well. SNe 2006X and 2020mvp were removed due to high host extinction, as suggested by the high \ebvhost values ($1.357$ and $0.792$\,mag, respectively) obtained with \snoopy fits to the \snia sample, using the \EBVmodel model\footnote{This model has \tmax, the distance modulus ($\mu$) and \ebvhost as free parameters.}. In particular, SN 2006X was already known to have high reddening \citep[e.g.,][]{Wang2008, Folatelli2010}. Another SN removed is 2020mbf, identified as a peculiar sub-type in this work. This object shows relatively flat NIR light curves, even in the \J band, with a very shallow valley, which hints toward a 06bt-like SN \citep{Foley2010}. The light curve and PISCOLA fit of this SN can be found in Fig.~\ref{fig:pec_SNeIa}. SN 2008hs is also identified as a peculiar sub-type and removed given that this object presents a secondary NIR peak happening at much earlier phase compared to the rest of the sample, hinting toward a possible  transitional\, \snia (see Fig.~\ref{fig:pec_SNeIa} as well). We note, however, that this SN has not been classified as such and it even passes the color-stretch cut applied below. Additionally, as PCA does not propagate uncertainties, we only selected SNe with relatively small photometric uncertainties ($\sigma_m$). We chose those with $\sigma_m < 0.2$\,mag across the entire phase range used (see below). Although applying a lower uncertainty threshold grants more precision, it also reduces the sample size, which is already very limited. We then used the color-stretch parameter, \sbv\footnote{This parameter measures the stretch in the $(B-V)$ color curve that peaks around $30$ days after optical maximum, combining the stretch and color parameters into a single one.} \citep{Burns2014} to remove the SNe falling outside the "normal"\, expected range (i.e., those with $0.6 \leq s_{BV} \leq 1.4)$. SNe 2007on, LSQ12fvl, and 2020uea did not pass this last cut, given their lower values. 

To summarize, from the initial sample of 158 \sneia, and after removing those with bad fits, peculiar shapes, relatively large uncertainties, and \sbv outside the adopted range, only 47, 36, and 25 SNe remained for the \textit{YJH} bands, respectively. The cuts applied in this section are presented in Table~\ref{tab:further_cuts}.

\begin{table}[!h]
\caption{Further cuts applied to the initial sample of 158 SNe~Ia (Section~\ref{subsec:preprocessing}).}
\centering

\setlength{\tabcolsep}{10pt}

\begin{tabular}{lccccccccc}
\hline
Cut & SNe per filter (Y,\textit{J},\textit{H}) \\
\hline
Before cuts$^a$             & 124,156,155 \\
Fit quality$^b$         &  116,148,147 \\
High \ebvhost           &  114,146,145 \\
Peculiar SNe            &  113,144,143 \\
$0.8\leq s_{BV}\leq1.2$ & 96,126,125 \\
$\sigma_m \leq0.2$\,mag & 47,36,25  \\
\hline
\end{tabular}

\begin{tablenotes}
 \item \textbf{Notes.} $^a$Not every SN has photometry on every NIR band. $^b$Qualitative cut from eye inspection of the fits.\\ 
\end{tablenotes}
\label{tab:further_cuts}

\end{table}

\begin{figure}[h!]
    \includegraphics[width=\columnwidth]{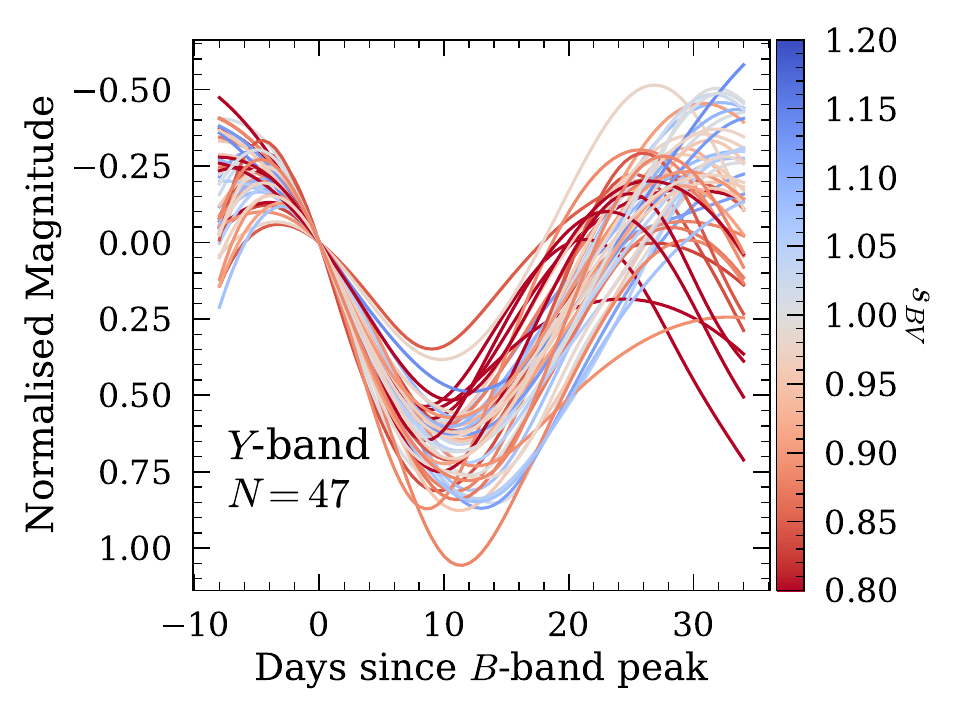}
    \includegraphics[width=\columnwidth]{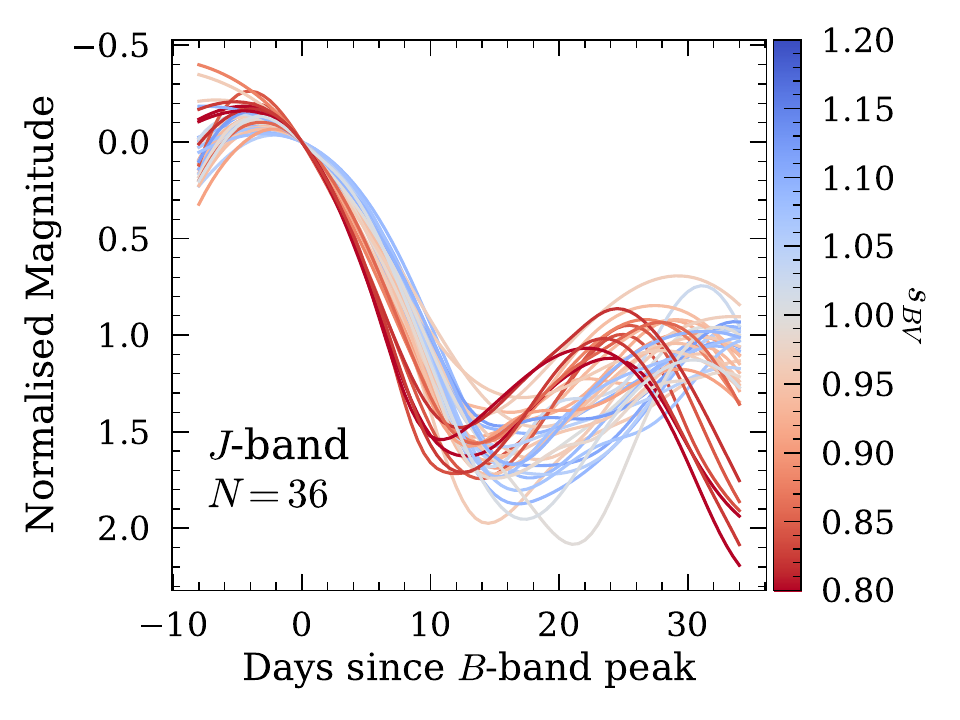}
    \includegraphics[width=\columnwidth]{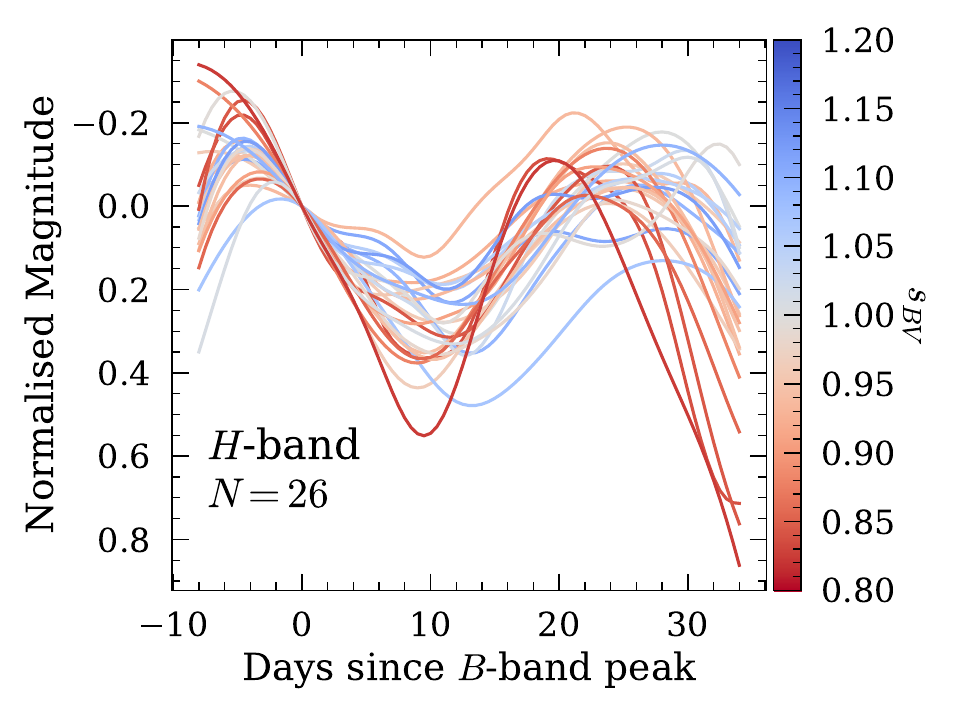}
    \caption{\sneia rest-frame light curves after the preprocessing steps of Sect.~\ref{subsec:preprocessing}. The phase range covers from $-8$ to $+34$ days with respect to optical peak. The light curves are color-coded by $s_{BV}$ value, obtained with \snoopy fits using the \maxmodel model, limiting the color range between $0.8 \leq s_{BV} \leq 1.2$ for visualization purposes, as the bulk of the sample falls within this range.}
    \label{fig:light_curves}
\end{figure}

After these cuts, the NIR light curves were shifted in time to match at the epoch of \textit{B}-band maximum luminosity, as obtained with \snoopy using the \maxmodel model. We then selected the light curve range between $-8$ and $+34$ days with respect to  the \textit{B}-band peak, for each band independently. This somewhat arbitrary range approximately covers the first NIR peak, which happens at $\sim5$\,days before optical peak and extends to the secondary peak, which occurs after $20$ days (see Fig.~\ref{fig:light_curves}; \citealt{Dhawan2015}). 

The last step of this process is normalizing the light curves by the brightness at the time of \textit{B}-band peak. This prevents the global scale from dominating the sample variance and allows the PCA decomposition to focus on other light curve features instead. Additionally, this removes the dependence on distance, which is not normally known with high accuracy for these extragalactic transients.

Figure~\ref{fig:light_curves} shows the \textit{YJH}-band light curves of \sneia after the aforementioned prepossessing steps. In the case of the \J and \H bands, around one third of the SNe come from CSP and one third from the literature, while for the \Y band, around $50\%$ come from CSP, one third from DEHVILS, and a small handful of objects  from the literature.
In general, a similar trend is observed for the \textit{YJH}-band light curves, where the SNe with lower \sbv values have earlier secondary peaks compared to those with higher values, following the same behavior described by \cite{Phillips2012} (see also Sect.~\ref{subsubsec:peak_abs}).

In terms of limitations and caveats, it is worth noting that the samples used are very limited in number. Given the requirements, the samples are mostly biased toward very nearby SNe ($z \lesssim 0.04$, with only a few SNe, mostly from DEHVILS, between $0.04 < z \lesssim 0.08$) and most likely bright events. However, at low redshifts $K$-corrections are expected to be small. Although the samples for the different bands have different number of SNe, we note that their $s_{BV}$ distributions are approximately consistent, with most SNe within the $ 0.8 < s_{BV} < 1.2$ range.

\subsection{Light-curve decomposition}
\label{subsec:decomposition}

\begin{figure*}[ht!]
    \includegraphics[width=\textwidth]{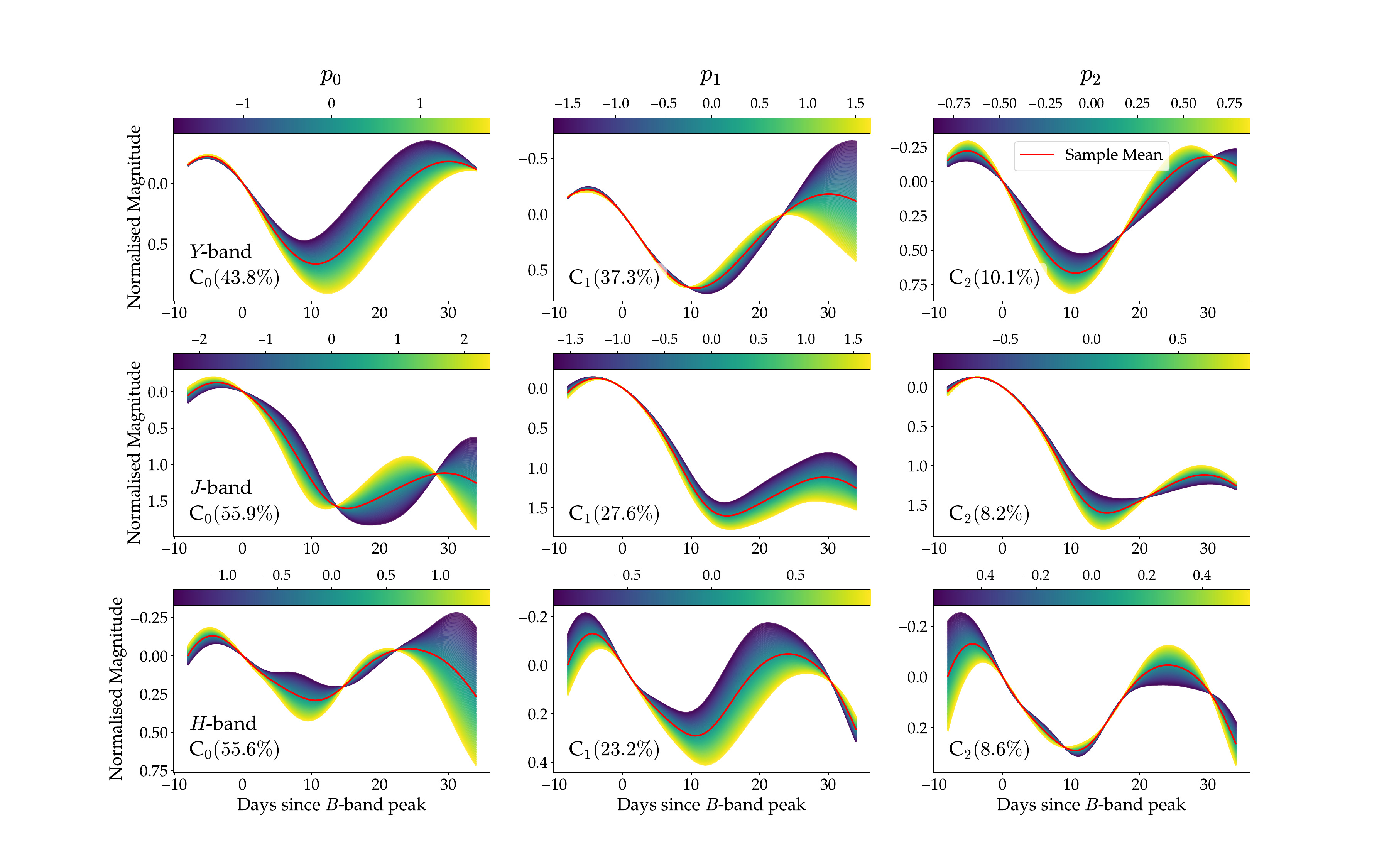}
    \caption{PCA decomposition for the \textit{Y}- (top row), \textit{J}- (middle row), and \textit{H}-band (bottom row) light curves. Color-coded are the contributions of each of the coefficients with a 2$\sigma$ range: $p_0$ (left column), $p_1$ (middle column), and $p_2$ (right column). The sample mean is shown as a solid red line. The values in parentheses are the percentages of the explained variance. Three components explain $\sim90\%$ of the variance for each of the bands.}
    \label{fig:lcs_decomposition}
\end{figure*}

In Fig.~\ref{fig:lcs_decomposition}, we show the \textit{YJH}-band light curves (top, middle, and bottom rows, respectively) decomposed into three PCA components. The first component (left column) captures most of the contribution to the light-curve variability ($\sim45-55\%$), while the second one (middle column) has moderate contribution ($\sim25-35\%$), and the third component (right column) mainly captures the remaining diversity ($10\%$), although it could also have contribution from noise within the data. In total, two PCA components explain around $80\%$ of the variability for each NIR band, while three components explain $90\%$, demonstrating their homogeneity.

Following the PCA decomposition, for each band $X$, the light curve can be described by

\begin{equation}
    LC^{\rm SN} = \overline{LC} +  \sum_i^{N=3} C_i \times p_i^{\rm SN},
\end{equation}

\noindent where $\overline{LC}$ is the light-curve mean for the sample (solid red line in Fig.~\ref{fig:lcs_decomposition}), and the three ($N=3$) components ($C_i$) are in common for the entire sample used, while the coefficients ($p_i$) depend on each SN. In the following section, we analyze what each of these components represents, how they contribute to the NIR light-curve shape, and we look for correlations with other parameters.

\section{Analysis}
\label{sec:analysis}

In the previous section, the NIR light curves were decomposed using PCA. As a reminder to the reader, the PCA components ($C_i$) are orthogonal, providing different but complementary information. This section provides a detailed discussion of these components.

\subsection{PCA components contribution to the NIR light curves}
\label{subsec:pca_comp}

As can be seen from Fig.~\ref{fig:lcs_decomposition}, the first PCA component ($C_0$) encompasses around half of the explained variance for each of the bands, while $C_0$ behaves in a similar way for the \J and \H bands; that is, it mainly describes how early or late the local minima and secondary peaks occur, with some contribution to the secondary peak brightness. For the \Y band, $C_0$ mainly controls the depth of the local minimum, including contribution to the timing of the secondary peak as well, but with a somewhat opposite effect compared to the other two bands. This difference between the bands might be surprising as we would expect the NIR light curve to behave in a similar manner. However, the physical processes that each wavelength range probes can differ (see more details below).

Examining the second PCA component ($C_1$) reveals its primary contribution to the brightness of the secondary peak across all bands, along with some influence on the timing of the secondary peak in the \Y and \H bands. However, it also controls the depth of the local minima for the \J and \H bands, and the brightness of the first peak for the \H band. It is also worth noting that the variability around first peak captured by the PCA on the \Y and \J bands is relatively small, but not so much for the \H band, which could be due to a wider range in the phase of the first NIR peak found at this latter band \citep{Dhawan2015}. We also find a moderate correlation between $C_1$ from the \Y band and $C_0$ from the \J and \H bands. Additionally, $C_1$ from the \H band is especially similar to $C_0$ from the \Y band (see Fig.~\ref{fig:lcs_decomposition}). This shows that the different components trace similar behavior for all the NIR bands, but with different relative importance.

As the third PCA component ($C_2$) only contributes to $\lesssim10\%$ of the variance in all bands and as it could be capturing noise in the data, we refrain from any interpretations and do not consider it in the rest of the analysis. To summarize, both $C_0$ and $C_1$ components mainly describe the timing and prominence of the secondary peak, but with somewhat opposite effects. Henceforth, we refer to the PCA components ($C_i$) and coefficients ($p_i$) interchangeably.

\subsection{Physical interpretation of the PCA components}
\label{subsec:pca_physics}

Interpreting PCA components presents a challenge due to the complexity of the underlying physics governing NIR light curves, where multiple physical processes are intertwined. \citetalias{Kasen2006} provided one of the few most recent detailed theoretical models of \sneia NIR light curves available, serving as the primary reference for this study. In particular, \citetalias{Kasen2006} used radiative transfer calculations, assuming local thermodynamic equilibrium (LTE), to calculate the optical and NIR light curves of \sneia, produced by Chandrasekhar-mass WDs. They also studied the effect of three main physical properties on the secondary NIR peak: the mixing of \nickel, the mass of \nickel (\mni), and the mass of stable iron-group elements (\mfe) produced by electron-capture elements and as a consequence of progenitor metallicity.
These properties are detailed below. 
\begin{itemize}
    
    \item Mixing: the mixing of iron-group elements, in particular \nickel, into the outer layers of the ejecta hastens the occurrence of the secondary peak by increasing the extent of the iron core. Thus, more mixing translates into an earlier secondary peak as the recombination of iron-group elements (\ion{Fe}{iii}$\rightarrow$\ion{Fe}{ii}) happens at earlier epochs.
    
    \item \mni: this parameter has two opposing effects. On the one hand, a larger \mni implies a larger iron core, hastening the appearance of the recombination of iron-group elements. On the other hand, a larger \mni implies higher temperatures, which means that the recombination temperature is reached at later epochs as it takes longer to cool down. \citetalias{Kasen2006} showed that the latter effect dominates over the former one.
    
    \item \mfe: although \snia explosions primarily produce radioactive \nickel, some non-negligible fraction of stable iron-group elements can be produced. The larger the amount of \mfe produced by electron-capture elements (e.g., $^{57}$Co, $^{58}$Ni, $^{54}$Fe), the larger the size of the iron core,  hastening the occurrence and increasing the prominence of the secondary peak. The progenitor metallicity also plays an important role in the production of stable iron-group elements. When \mfe is produced at expense of \mni, lower temperatures are reached, producing an earlier and somewhat fainter secondary peak.
 
\end{itemize}

The variability captured by $C_0$ for the \J and \H bands is dominated by the timing of the secondary peak, with some degree of dependence on its brightness (see Fig.~\ref{fig:lcs_decomposition}). From the physical properties studied by \citetalias{Kasen2006}, \mni and metallicity produce a similar behavior as observed for this component, where \sneia with later secondary peaks are also brighter. \cite{Ashall2019a} and \cite{Ashall2019b} showed that the blue-edge velocity of the Fe/Co/Ni emission feature in the \H-band, measured between a week or two after optical maximum, traces the outer \nickel layer, which could potentially be used as a way of disentangling the effects of \mni and metallicity. However, to see whether one effect does indeed dominate over the other, a more detailed analysis of physical properties is needed (see Sect.~\ref{subsec:comparison}).

In the \J band, $C_1$ primarily captures to the relative brightness (see Fig.~\ref{fig:lcs_decomposition}), with a very small effect on the timing of the secondary peak. The physical properties studied by \citetalias{Kasen2006} predominantly affect the timing of the secondary peak, so in order to reproduce the observed behavior, a combination of these with opposing effects on the timing would be required. For instance, a mixture of \mni and \mfe produced from electron-capture elements, or \nickel mixing, would affect the brightness and leave the timing of the secondary peak relatively unaffected. For the \H-band $C_1$ component (see Fig.~\ref{fig:lcs_decomposition}), both the \nickel mixing and \mfe produced from electron-capture elements could explain the observed behavior; especially the latter property, which has a greater effect on the secondary peak brightness.

Although \citetalias{Kasen2006} did not model the \Y-band light curve, the analysis of both the \textit{I} and \J bands can be used as an approximation. The \Y-band $C_0$ component resembles the $C_1$ component of the \H band, and vice-versa (see Fig.~\ref{fig:lcs_decomposition}), which would suggest that the same physical properties dominate these components. In other words, the same physical properties behind $C_1$ and $C_0$ for the \H band could explain the behavior of $C_0$ and $C_1$, for the \Y band. Certain physical properties, such as opacity and energy distribution, influence the SEDs of \sneia in varying ways across different wavelengths, possibly explaining why some physical properties would be dominant in one band and some in others. In addition, different NIR wavelengths probe different layers within the ejecta (and more deeply than the optical) at the same epoch \citep{Wheeler1998}, which is also a viable explanation.

The analysis of spectral features can also be used to help explain the difference in variability between different wavelength ranges. As spectroscopic observations at NIR wavelengths are relatively limited, theoretical models provide an excellent alternative. \cite{Collins2022} presented 3D radiative transfer simulations of double-detonation sub-Chandrasekhar-mass WDs as a plausible scenario of \snia explosions. These simulations show that the NIR spectra of \sneia around the time of secondary NIR peak are dominated by iron-group elements\footnote{Christine Collins (private communication).}. This is in good agreement with previous theoretical works (e.g., \citetalias{Kasen2006}) and observations \citep[e.g.,][]{Marion2003, Marion2009}. In particular, the \Y band ($\lambda \sim1.0\,\mu$m) is dominated by \ion{Fe}{II}, followed by \ion{Co}{II}; \H band ($\lambda \sim1.6\,\mu$m) is dominated by \ion{Co}{II}, followed by \ion{Fe}{II}; while the \J band ($\lambda \sim1.2\,\mu$m) has a relatively similar contribution from \ion{Fe}{II}, \ion{Co}{II,} and \ion{Ca}{II}. This is in agreement with observations, except for the \J band, where not much \ion{Co}{II} is observed and where \ion{Ca}{II} is just observed around the optical peak \citep{Marion2003, Marion2009}. Nonetheless, this shows that despite their apparent similarity, the NIR bands have intrinsic differences as well, thus probing different aspects of the explosion. The opposite relative contribution from \ion{Fe}{II} and \ion{Co}{II} to the \Y and \H bands could possibly explain the relation between the $C_0$ and $C_1$ components seen for these bands (see Fig.~\ref{fig:lcs_decomposition}). However, one caveat related these simulations is that they are not fully non-LTE \citep{Collins2022}, which tends to produce less realistic results. 

To summarize, the $JH$-band $C_0$ and $Y$-band $C_1$ components seem to mainly trace the effect of \mni and metallicity on the secondary NIR peak, while the $JH$-band $C_1$ and $Y$-band $C_0$ components could capture the effect of \nickel mixing and \mfe. Additionally, some of the differences between the bands may be attributed to the different ejecta layers they trace. In the next section, we  explore correlations between light-curve parameters and PCA components to help disentangle the effects of different physical properties.

\subsection{Comparison to light-curve parameters}
\label{subsec:comparison}

To have a better understanding about the physics behind each of the PCA components, we proceeded to compare them against different light-curve parameters. We calculated the linear relation for each pair of parameters, by using the Pearson correlation coefficient ($\rho$) and $p$-value. The values of $\rho$ range from $-1$, completely anti-correlated, to $1$, completely correlated, where $0$ is no correlation. The $p$-value gives the probability of the null hypothesis (slope of the linear relation being equal to zero) being true. Commonly, if the $p$-value is below $0.05$, the null hypothesis is discarded. We used Monte-Carlo sampling (1000 realizations) from the light-curve parameters and their respective uncertainties to propagate the errors into $\rho$ and the $p$-value. We assume no uncertainty for the PCA coefficients ($p_0$ and $p_1$) as PCA does not propagate the uncertainty.

\subsubsection{NIR peak absolute magnitude}
\label{subsubsec:nir_peak_mag}

\begin{figure*}[h]
    \includegraphics[width=\textwidth]{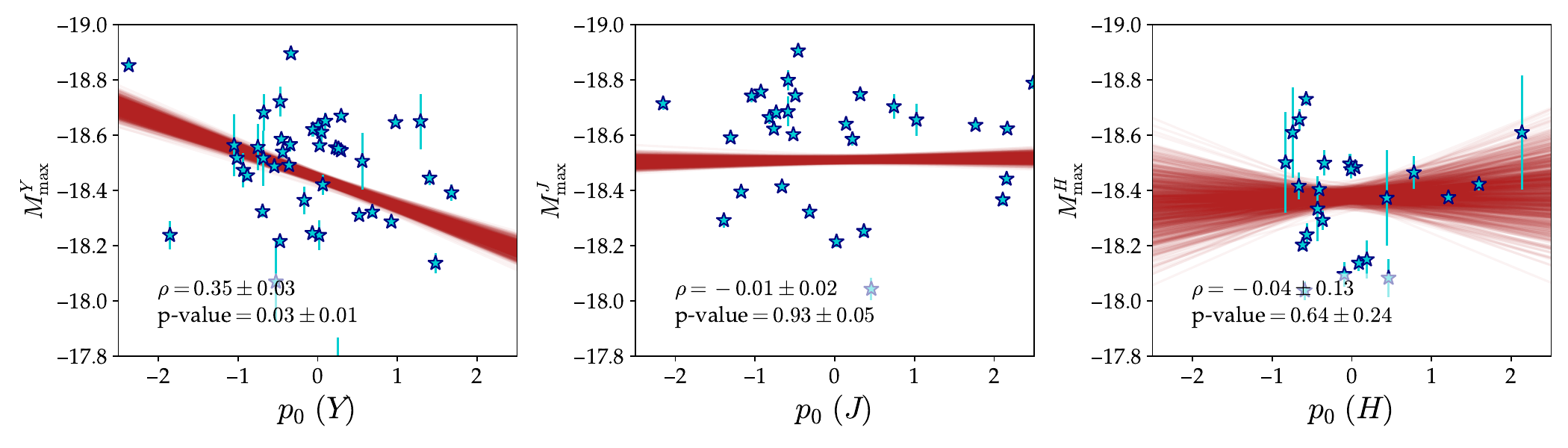}
    \includegraphics[width=\textwidth]{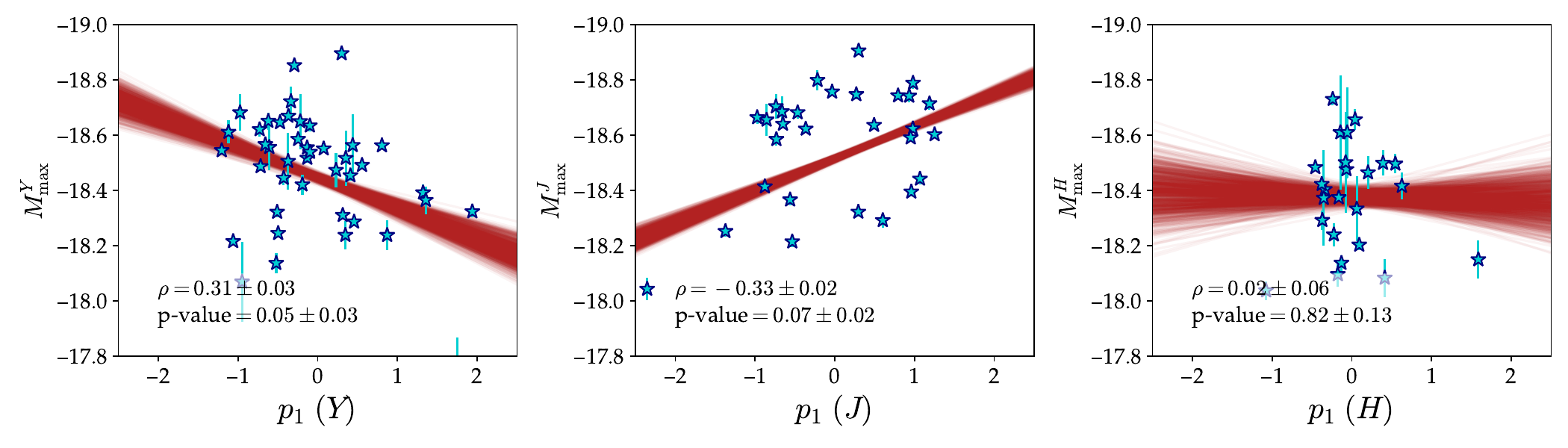}
    \caption{Comparison of the first ($p_0$; top row) and second ($p_1$; bottom row) PCA coefficients vs. the peak absolute magnitude for each NIR band. The Pearson correlation coefficient ($\rho$) and $p$-value for each of the comparisons, with their respective 1$\sigma$ uncertainty estimated by Monte-Carlo sampling, are shown with the respective linear relations (red lines). The null-hypothesis is that there is no correlation (zero slope). The panels have the same \textit{x}- and \textit{y}-axis ranges for visualization purposes.
    }
    \label{fig:Mmax_vs_p01}
\end{figure*}

In the top and bottom panels of Fig.~\ref{fig:Mmax_vs_p01}, relations between the NIR peak absolute magnitudes\footnote{Assuming a cosmology with $\Omega_m = 0.3$, $\Omega_{\lambda} = 0.7$ and $H_0 = 70$\,\hunits.} and the first two PCA components ($p_0$ and $p_1$) are presented. No correlations are found for the \J\footnote{The correlation for $p_1$ in the \J band is driven by a single data point, at the bottom left.} and \H bands, although the \Y band shows a significant but relatively weak correlation. If \sneia were to be standardizable candles in the NIR, correlations could be expected between their peak brightness and other light-curve parameters, which is not what the \J and \H band show. However, the \Y band, which closer to the optical wavelengths than the other two bands, does show some correlation. The standardization of the \Y band is studied in Sect.~\ref{subsec:standardization}.

\subsubsection{Color-stretch, \sbv}
\label{subsubsec:peak_abs}

\begin{figure*}[h]
    \includegraphics[width=\textwidth]{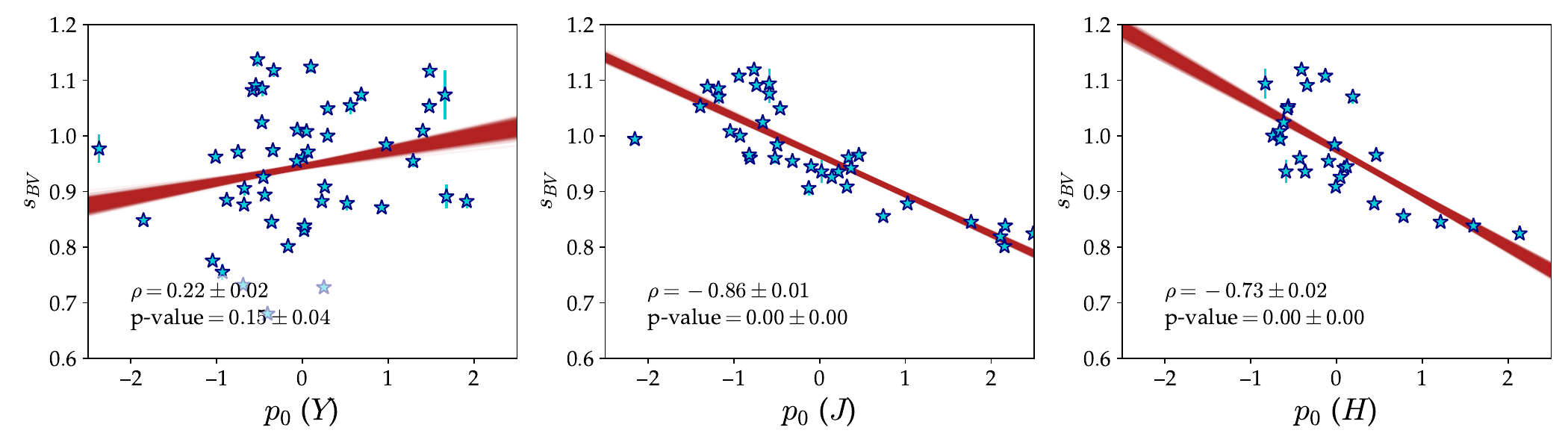}
    \includegraphics[width=\textwidth]{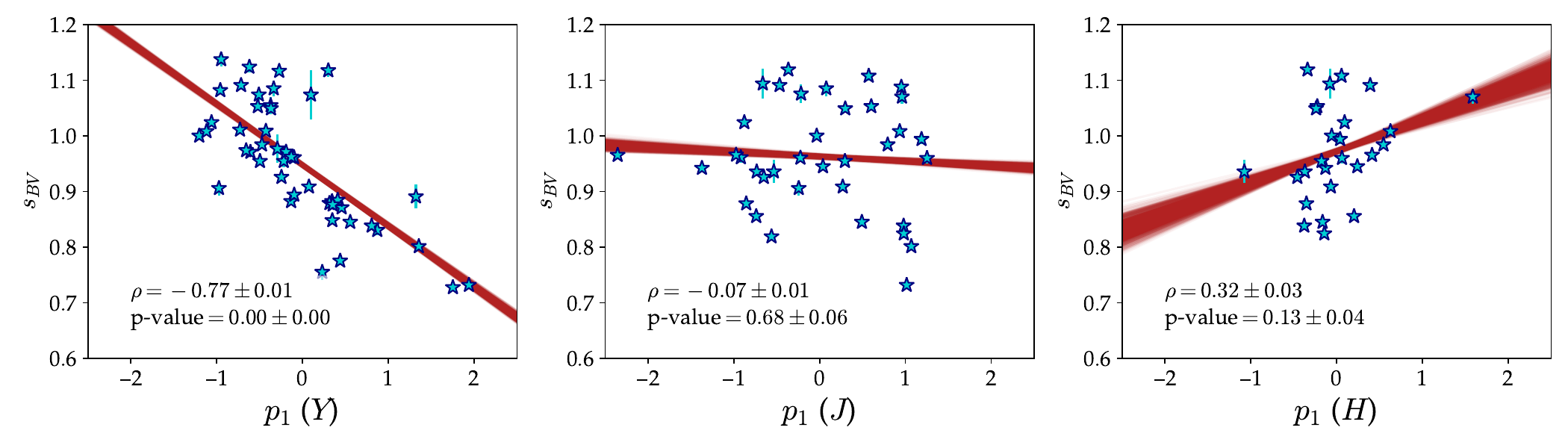}
    \caption{Same as Fig.~\ref{fig:Mmax_vs_p01}, but for $s_{BV}$, obtained with \snoopy using the \maxmodel.
    }
    \label{fig:stretch_vs_p01}
\end{figure*}

In the top panels of Fig.~\ref{fig:stretch_vs_p01}, we show the relation between $p_0$ and \sbv for each of the NIR bands. \sbv works as a tracer of the time when the iron-group elements recombine (\ion{Fe}{iii}$\rightarrow$\ion{Fe}{ii}; appendix of \citealt{Galbany2023}). If this were true, we would expect very strong correlations between these parameters. Indeed, it can be seen that both \J and \H bands show strong correlations between \sbv and $p_0$, but the \Y band does not. However, the latter shows a strong correlation between \sbv and $p_1$, while the other two bands do not (see bottom panels of Fig.~\ref{fig:stretch_vs_p01}). As \sbv correlates with \mni, this is evidence in favor of our interpretation in Sect.~\ref{subsec:pca_physics}. Additionally, \cite{Lu2023} already showed that \sbv can successfully capture the NIR light-curve shape of \sneia. However, it is worth noting that this parameter is not correlated with all PCA components, so the NIR variability may be more complex than we could expect. One caveat in this comparison is that the \snoopy NIR templates, used for correcting the light curves (Sect.~\ref{sec:rest_lcs}), are based on the \sbv parameter; thus, a completely independent comparison is not possible as the PCA and \sbv are linked from the start. 

\subsubsection{Optical peak absolute magnitude}
\label{subsubsec:opt_peak_mag}

\begin{figure}[h]
    \includegraphics[width=\columnwidth]{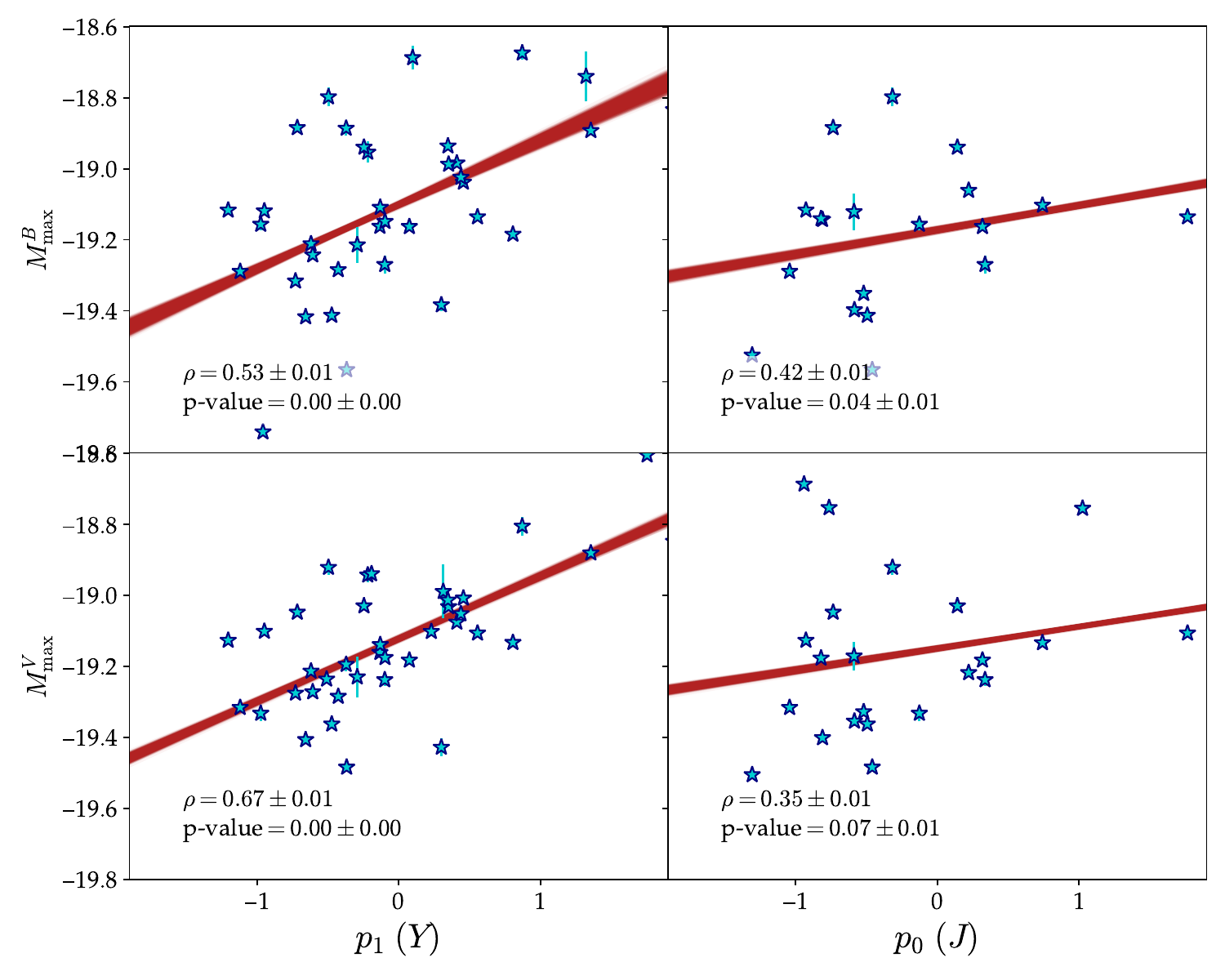}
    \caption{Comparison between \textit{B}- (top panels) and \textit{V}-band (bottom panels) peak absolute magnitudes versus $p_1$ for the \Y band (left panels) and $p_0$ for the \J band (right panels).}
    \label{fig:Mopt_vs_p10}
\end{figure}

It is known that the light curves of \sneia are primarily powered by the decay of \nickel \citep{Colgate1969}. Hence, their peak bolometric brightness is proportional to the \mni produced in the explosion (known as Arnett's rule; \citealt{Arnett1982}). As most of the luminosity ($\sim85\%$) of \sneia is emitted at optical wavelengths \citep{Maguire2017}, the optical luminosity at its peak can be used as a proxy for \mni \citep[e.g.,][]{Stritzinger2006}. We found a correlation between the optical peak absolute magnitudes (\textit{BV} bands) and $p_1$ for the \Y band and $p_0$ for the \J band. This is shown in Fig.~\ref{fig:Mopt_vs_p10}. These correlation aligns with our analysis from Sect.~\ref{subsec:pca_physics}, where $C_1$ describes the contribution of \mni for the \Y band and $C_0$ for the \J band, although no correlation was found for the \H band. It is possible that \mni could have less influence toward redder wavelengths given the diminishing strength and significance of the correlation from \Y to \H.

\subsection{Comparison to host-galaxy properties}
\label{subsec:hosts}

It has been known for over a decade that the environment in which \sneia reside affects their peak optical brightness, where SNe in more massive environments are brighter, on average, compared to those in less massive hosts, after standardization \citep[e.g.,][]{Kelly2010, Lampeitl2010, Sullivan2010}. However, whether this dependence is mainly due to external factors, such as dust \citep{BS21}, or other intrinsic effects \citep[e.g.,][]{Ginolin2025}, is still a matter of debate. 

Overall, NIR observations of \sneia can provide clues on this matter as these wavelengths are less affected by dust extinction. In fact, it has already been shown in recent years that the NIR peak brightness of \sneia does depend on the environment in which they reside \citep[e.g.,][]{Uddin2020, Ponder2021, Uddin2024}, most likely implying that dust is not the dominant mechanism behind this relation. However, to our knowledge, the effect of the environment on the NIR light curves beyond first peak has not been explored yet and this approach could provide new insights into the physics of these transients.

In this work, we used previous estimates of the host-galaxy masses from the literature. \citet{Uddin2024} provides host-galaxy masses for the entire CSP sample (\cspi and \cspii), which represents a large fraction of the samples used (see Sect.~\ref{subsec:preprocessing}). In addition, \citet{Ponder2021} estimated host-galaxy masses for several SNe from different surveys and others from the literature. Although they also included \cspi, we prioritized the values from \citet{Uddin2024} for consistency. Finally, \citet{Peterson2024} provided values for the DEHVILS sample.\footnote{Erik Peterson was kind enough to share with us the host-galaxy masses, including uncertainties, estimated with Prospector \citep{Leja2017, Johnson2021}, as those in \citet{Peterson2024} do not include these uncertainties.} Although different studies estimate host masses with different methods, they usually agree relatively well.

\begin{figure*}[h]
    \includegraphics[width=\textwidth]{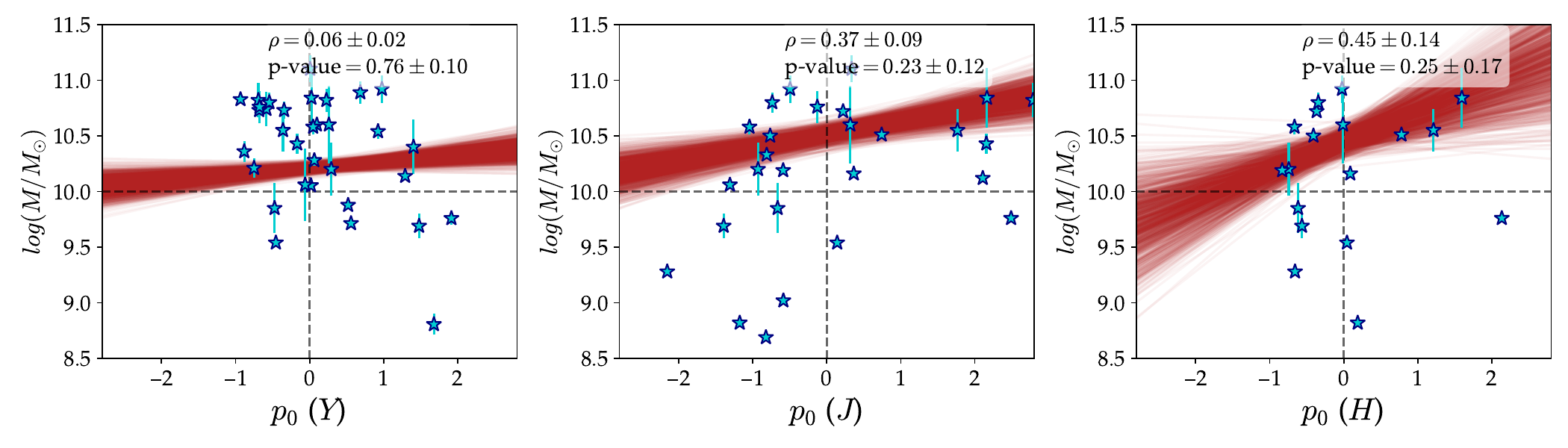}
    \caption{Comparison between host-galaxy stellar mass, \hostmass and $p_0$. The vertical dashed line marks $p_0=0.0$, dividing the \snia sample at the mean, while the horizontal lines mark \hostmass$= 10$, commonly used for splitting \sneia samples into high- and low-mass hosts.}
    \label{fig:hostmass_vs_p0}
\end{figure*}

In Fig.~\ref{fig:hostmass_vs_p0}, the relation between host-galaxy stellar mass, \hostmass, and $p_0$ is shown. Although no correlation is found, a possible lack of SNe with faint \J-band secondary peak in low-mass hosts is apparent. However, this could be due to a bias given the relatively small number of SNe in the sample. Additionally, no clear trend was found with $p_1$, possibly implying that the environment has no visible effect on the secondary NIR peak of \sneia; however, no strong conclusions can be drawn. We do note that the host-galaxy mass, while easier to estimate than other host properties, might not adequately capture  the effect of the environment on \sneia NIR light curves. 

The relation between host-galaxy color excess, \ebvhost (obtained with \snoopy using the \EBVmodel), and the first two PCA components has been here studied as well. As with \hostmass, we did not find any clear correlation. Assuming that dustier galaxies are expected to be more metal-rich (i.e., with more metal-rich progenitors), some correlation would be expected given our analysis in Sect.~\ref{subsec:pca_physics}, but a more precise estimation of metallicity would be needed to draw any strong conclusions. Future works will extend to the use of other galaxy properties, such as the star formation rate and metallicity, to further study the impact of the environment on the secondary NIR peak of \sneia.

\subsection{NIR color-curves decomposition}
\label{subsec:color_decomposition}

\begin{figure}[h!]
    \includegraphics[width=\columnwidth]{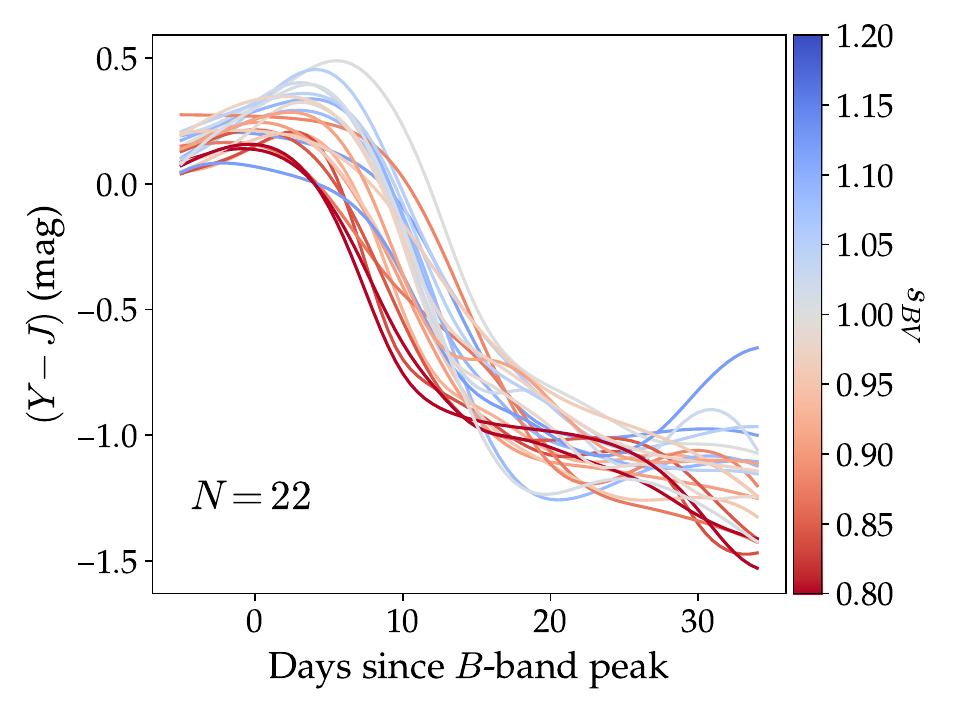}
    \includegraphics[width=\columnwidth]{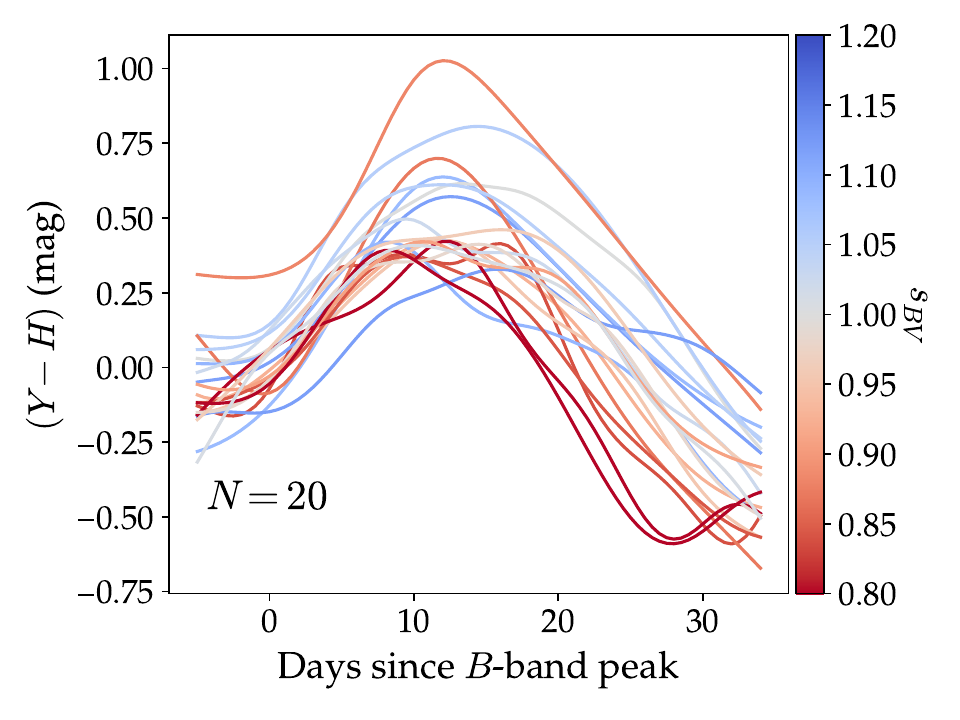}
    \includegraphics[width=\columnwidth]{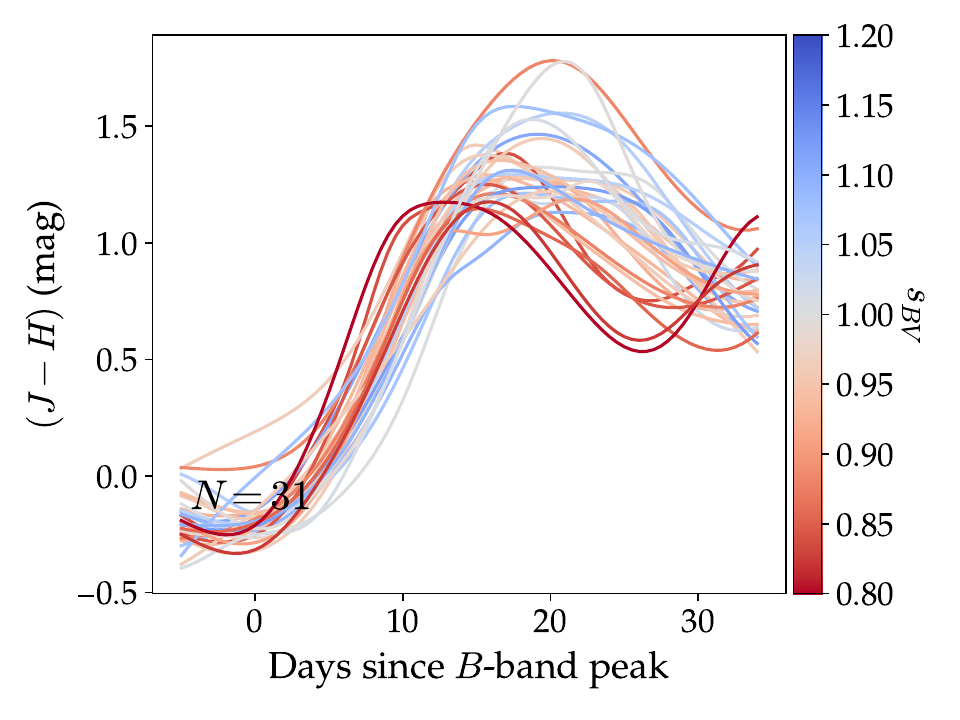}
    \caption{Same as Fig.~\ref{fig:light_curves}, but for the color-curves: \textit{(Y-J)} (top panel), \textit{(Y-H)} (middle panel), and \textit{(J-H)} (bottom panel).}
    \label{fig:color_curves}
\end{figure}

Following the PCA decomposition of the NIR light curves of \sneia, we studied the decomposition of the rest-frame, \textit{K}- and MW-corrected, NIR color curves: $(Y-J)$, $(Y-H)$, and $(J-H)$. The same criteria applied in Sect.~\ref{subsec:preprocessing} were applied here, except for the normalization in the \textit{y}-axis. The color curves are shown in Fig.~\ref{fig:color_curves}. We note that the number of \sneia used in this case is reduced compared to those in Sect.~\ref{sec:lc_decomposition}, due to the requirements of having two bands with proper coverage instead of one. An inspection of Fig.~\ref{fig:color_curves} reveals that the $(Y-J)$ and $(J-H)$ color curves follow a very similar evolution to those shown by \cite{Lu2023}, where there is a clear separation between \sneia with high and low \sbv. This means that \sneia that are optically bluer around optical maximum have on average redder NIR colors around the time of secondary NIR peak. However, the NIR color as a function of \sbv does not present a monotonic transition as smooth as that seen in \cite{Lu2023}.

\begin{figure*}[ht!]   \includegraphics[width=\textwidth]{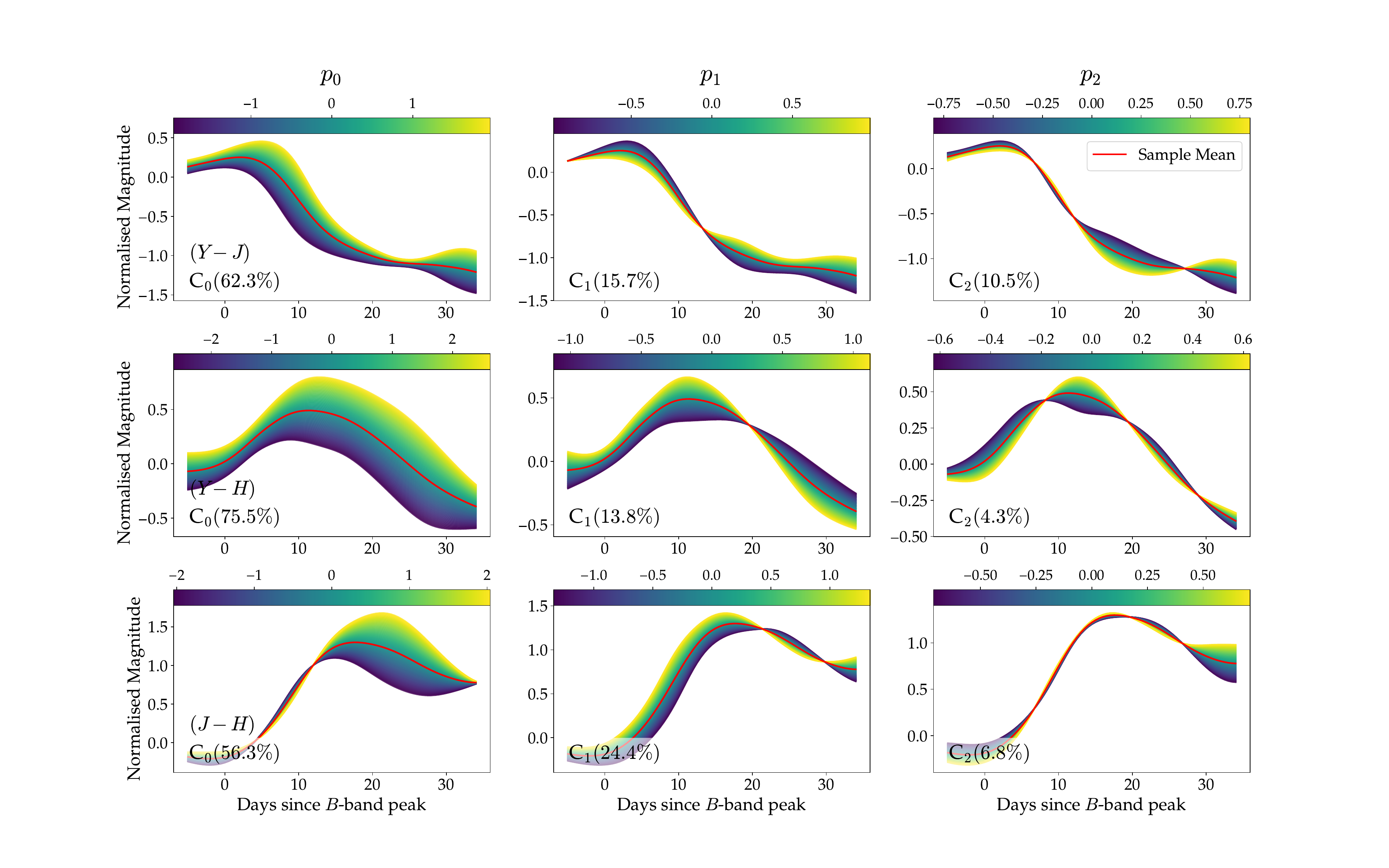}
    \caption{Same as Fig.~\ref{fig:lcs_decomposition}, but for PCA decomposition of the \sneia color curves: \textit{(Y-J)} (top row), \textit{(Y-H)} (middle row), and \textit{(J-H)} (bottom row).}
    \label{fig:color_decomposition}
\end{figure*}

The color-curve decomposition is presented in Fig.~\ref{fig:color_decomposition}. It can be noted that the first component ($C_0$) captures over $50\%$ of the variance for all color curves, while relatively low variance is captured by the other components. Additionally, it can be seen that NIR color curves present relatively high intrinsic uniformity around optical peak, also shown in Fig.~\ref{fig:color_curves}. 

The $C_0$ components mainly capture how blue or red the NIR color curves are, where SNe with higher NIR colors peak at later epochs (see Fig.~\ref{fig:color_decomposition}). Additionally, \sbv seems to trace the same behavior as $C_0$ (see Fig.~\ref{fig:color_curves}). It can also be seen that the NIR color curves of \sneia show some similarity in evolution with the $(B-V)$ curve, but reach their reddest point at much earlier phases: around $5$\,days for $(Y-J)$, $12$\,days for $(Y-H)$, and $18$\,days for $(J-H)$. In other words, $p_0$ could be linked to an \sbv-like parameter in the NIR and trace similar information. For instance, the peak in the $(J-H)$ curve approximately coincides with the local minimum on the \J band (see Figs.~\ref{fig:lcs_decomposition} and \ref{fig:color_decomposition}), possibly tracing the onset of the recombination of iron-group elements.

When comparing the PCA components against NIR peak absolute magnitude, we see no clear correlations, except for the $(Y-H)$ $C_1$ components. However, the variance explained by this component is relatively low ($\sim$14\%), so it does not carry much information. This lack of correlation is not surprising as similar results were found in general for the decomposition in Sect.~\ref{subsubsec:nir_peak_mag}.

\begin{figure*}[h!]
    \includegraphics[width=\textwidth]{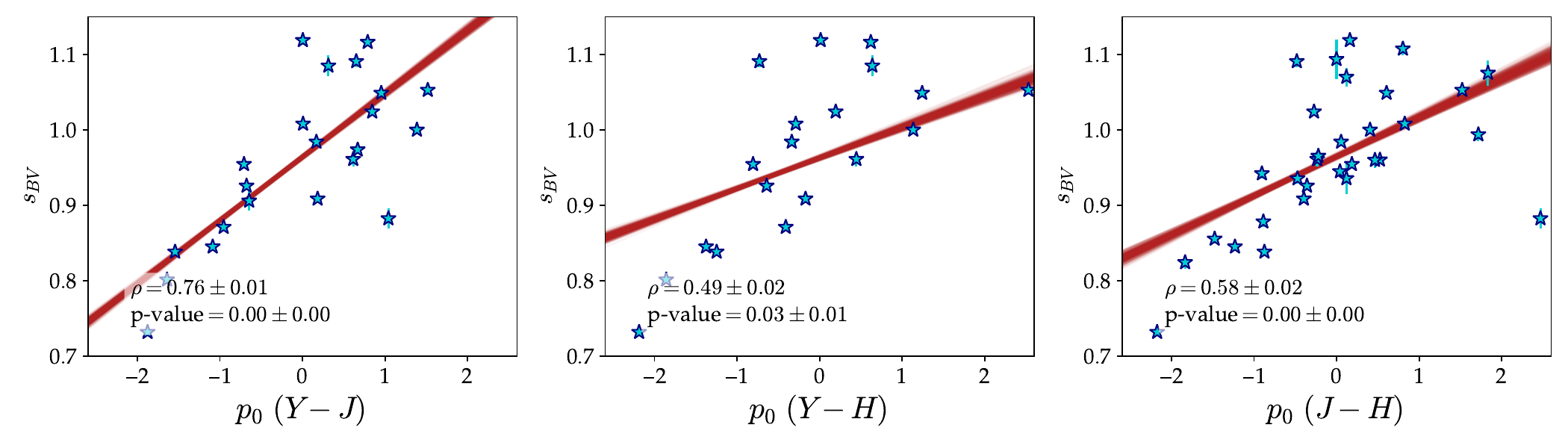}
    \caption{\sbv vs. PCA $p_0$ coefficient for the NIR color curves.}
    \label{fig:sbv_vs_p0_color}
\end{figure*}

In Fig.~\ref{fig:sbv_vs_p0_color}, we see  a clear correlation between \sbv and $p_0$ for all the NIR colors. Again, this is not surprising given  the NIR templates from \cite{Lu2023} were built as a function of \sbv. However, the scatter shows the complexity of  NIR variability, which cannot be entirely captured by a single parameter.

\begin{figure}[h!]
    \includegraphics[width=0.85\columnwidth]{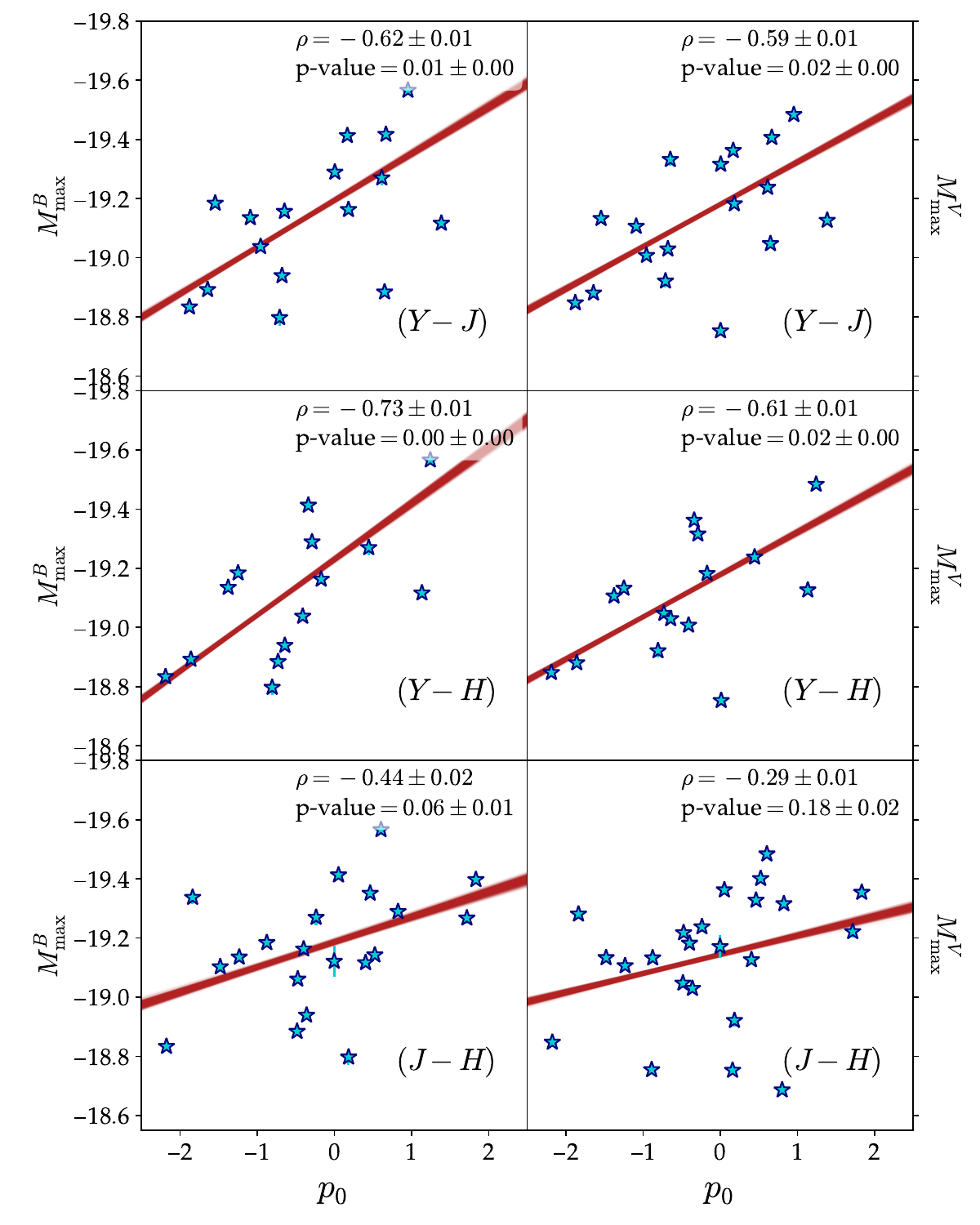}
    \caption{\textit{B}-band (left panels) and \textit{V}-band (right panels) peak absolute magnitude vs. PCA $p_0$ coefficient for the NIR color curves.}
    \label{fig:Mopt_vs_p0_colors}
\end{figure}

In the case of the peak absolute magnitude at optical wavelengths (both \textit{B} and \textit{V} bands), we see correlations with $p_0$ for all the NIR color curves (see Fig.~\ref{fig:Mopt_vs_p0_colors}). This is in line with the correlations found with \sbv and shows that \sneia with more \mni have redder NIR colors.

\begin{figure*}[h!]
    \includegraphics[width=\textwidth]{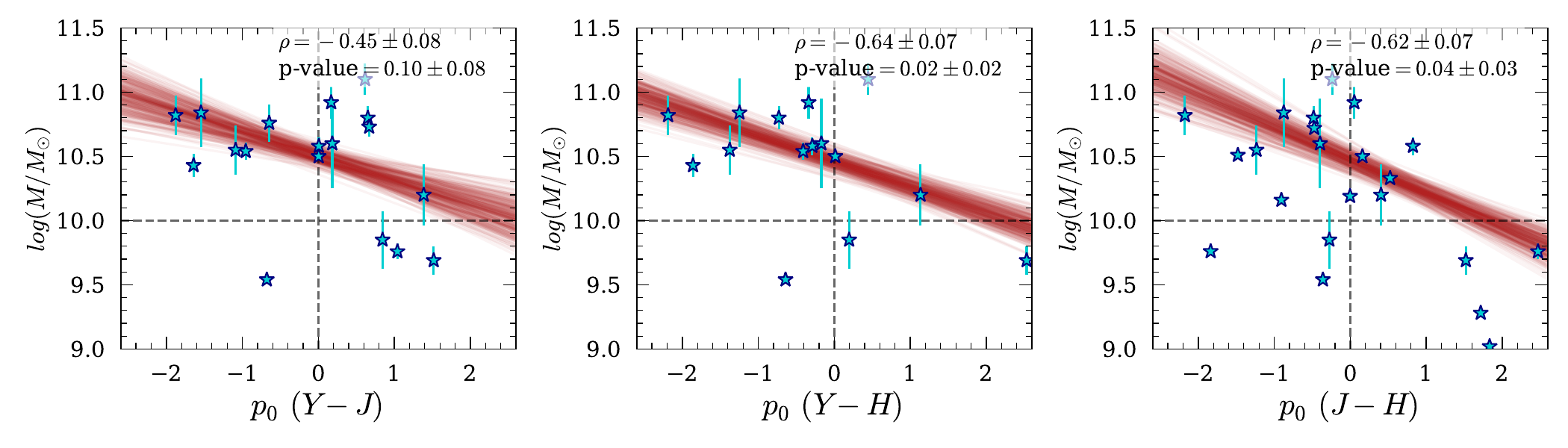}
    \includegraphics[width=\textwidth]{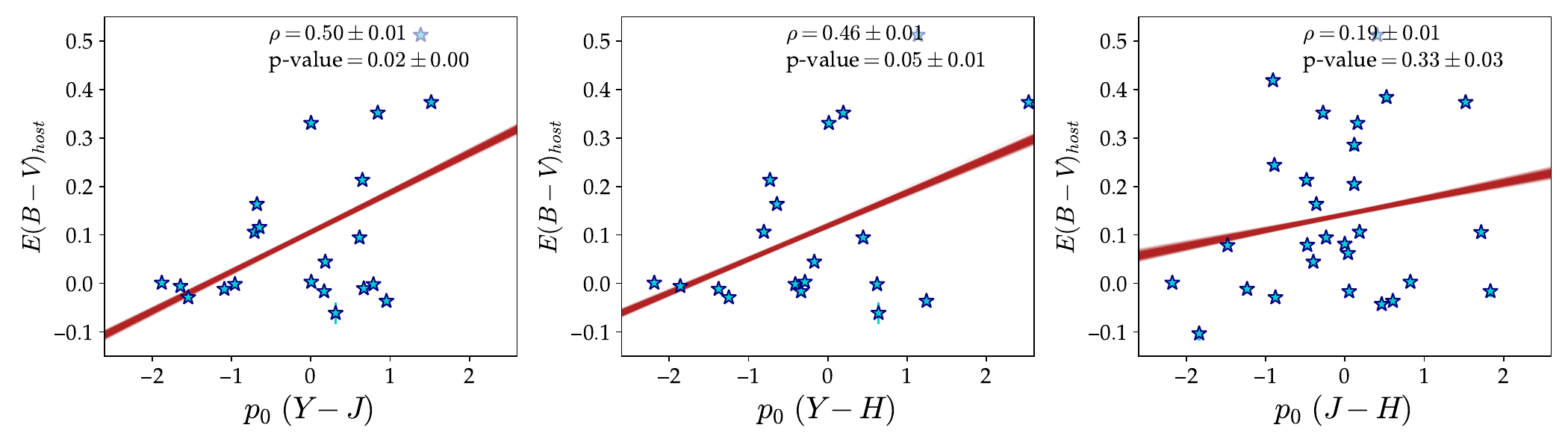}
    \caption{Comparison between \hostmass (top panels) and \ebvhost (bottom panels) vs. $p_0$ for the NIR colors.}
    \label{fig:host_vs_p0_color}
\end{figure*}

The comparison against host properties presents some interesting results as well. In the top panel of Fig.~\ref{fig:host_vs_p0_color}, the correlation between \hostmass and $p_0$ for the different color curves is shown. Although these correlations can be driven by a bias due to the small sample of \sneia, the same trend is observed for all color curves, where the redder objects are preferentially found in less massive galaxies (see Fig.~\ref{fig:host_vs_p0_color}). If this were true (and assuming that more massive galaxies are also more metal rich), this results would align with the expected effect of metallicity on the secondary NIR peak (fig.~13 in \citetalias{Kasen2006}), where the more metal-poor objects would have redder NIR colors and vice versa. However, a larger sample would be needed to confirm this.

When the same comparison is made against \ebvhost (bottom panel of Fig.~\ref{fig:host_vs_p0_color}), we see correlations for the $C_0$ component of $(Y-J)$ and $(Y-H)$. These are mainly driven by a couple of SNe (top right points in the bottom left and middle plots of Fig.~\ref{fig:host_vs_p0_color}): 2007ca and 2008fp, both from the CSP survey. A close inspection to the light-curve fits does not reveal any abnormality. When comparing to another SN with similar $p_0$ but lower \ebvhost, SN 2017cbv, these SNe show higher (i.e., redder) $(B-V)\sim$, $(Y_{\rm max}-J_{\rm max})$ and ($B_{\rm max}-J_{\rm max}$) values. When comparing to a SN with a similarly high \ebvhost but lower $p_0$, 2007le, the only difference is that the latter has a bluer $B_{\rm max}-J_{\rm max}$ value. Whether there is a potential bias in the estimation of \ebvhost is hard to tell, but the comparison between SNe, coupled with the fact that no correlation is seen between $C_0$ and $(J-H)$, possibly suggests that the correlations found might be driven by the relatively low statistics.

As \citetalias{Kasen2006} did not provide models for the color curves and no model for the \Y band, it is challenging to obtain more insights into the physical interpretation of these PCA components.

\subsection{NIR light-curve standardization}
\label{subsec:standardization}

Following the correlation found between \Y-band peak absolute magnitude ($M^Y$) and the PCA components (see Sect.~\ref{subsubsec:nir_peak_mag}), we proceeded to test the standardization of the \Y-band light curves of \sneia by adding correction parameters for the estimation of distances, expressed as 

\begin{equation}
    \mu_{\rm SN} = (m^Y + \alpha\times p_0 + \beta\times p_1) - M^Y ,
\end{equation}

\noindent where $m^Y$ is the \Y-band peak apparent magnitude, and $\alpha$ and $\beta$ are nuisance parameters. We followed a similar procedure as in \cite{Galbany2023}, but fixed $H_0 = 70$\,\hunits and fit for $M$ and $\sigma_{\rm int}$ (SN intrinsic scatter), with the addition of $\alpha$ and $\beta$. Since $H_0$ is fixed, distances from the calibrating SNe are not needed to anchor the $y$-axis in the Hubble diagram to break the degeneracy between $H_0$ and $M$.

\begin{figure}
    \includegraphics[width=\columnwidth]{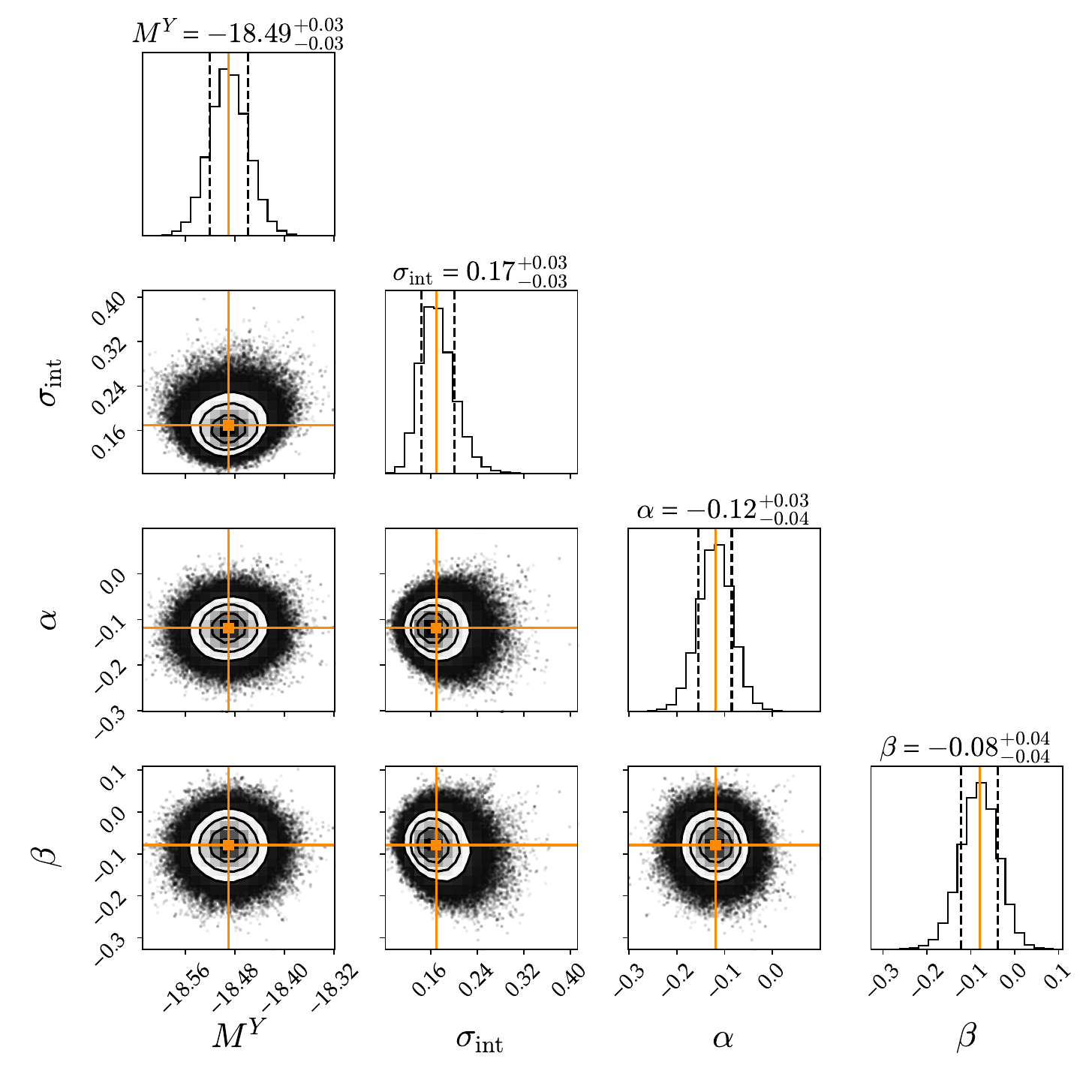}
    \caption{Corner plot of the MCMC posterior distributions used for building a \Y-band Hubble diagram. $\sigma_{\rm int}$ is the intrinsic scatter, i.e., leftover scatter unexplained by the standardization. The solid lines show the mean of the distributions while the dashed lines show the 16th and 84th percentiles ($1\sigma$).}
    \label{fig:Y_contours}
\end{figure}

\begin{figure}
    \includegraphics[width=\columnwidth]{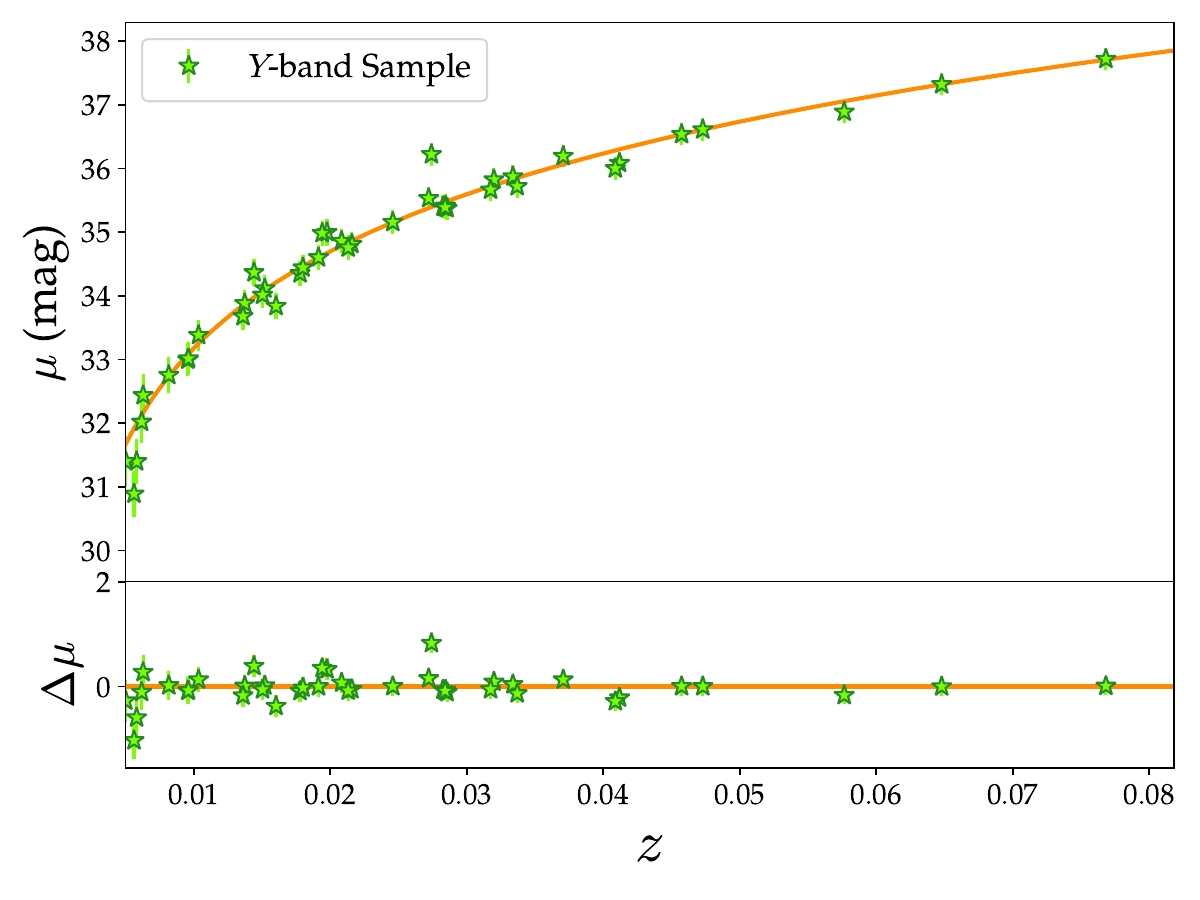}
    \caption{\Y-band Hubble diagram with the standardization from Fig.~\ref{fig:Y_contours}. The solid line represents the fitted cosmology, a flat $\Lambda$CDM cosmological model with $\Omega_{\rm m}=0.3$ and $H_0 = 70$\,\hunits. The Hubble residual presents a RMS of $0.27$\,mag and a normalised median absolute deviation of $0.10$\,mag.}
    \label{fig:Y_HD}
\end{figure}

Figure~\ref{fig:Y_contours} shows the corner plot of the Markov chain Monte Carlo (MCMC) posterior distributions of the fitted parameters used for building the \Y-band Hubble diagram with standardization (Fig.~\ref{fig:Y_HD}). The addition of standardization parameters is favored by around $3\sigma$ and $2\sigma$ significance for $p_0$ and $p_1$, respectively. Additionally, by correcting the light curves, $\sigma_{\rm int}$ is reduced from $0.21$\,mag to $0.17$\,mag. Although the quoted values might seem high at first, this is mainly attributed to the lack of uber-calibration of the photometry required for a precise cosmological analysis, which is beyond  the scope of this work\footnote{We note that restricting the sample to a single survey, such as CSP, largely reduces the scatter, but limits the sample size.}. The peculiar velocities of low-$z$ \sneia also affect the measured scatter, but removing these objects would further reduce the sample size.

While the standardization in the \Y band is favored, the \J and \H bands do not present any improvements with the addition of other parameters, reflecting their uniformity. The standardization using the PCA components from the NIR color curves is covered in Appendix~\ref{app:color_cosmo}.

\section{Summary and conclusions}
\label{sec:conclusions}

In this work, we  gathered a sample 
of 47, 36, and 25 \sneia with well-sampled \textit{YJH}-band light curves, respectively, to analyze their variability with PCA, a common machine-learning technique.
SNe~Ia were selected from various sources, including the CSP, CfAIR2, RATIR, and DEHVILS surveys, as well as some objects drawn from the literature. Each selected object has at least five NIR photometric epochs in any band, with coverage spanning from the first to the second peak. The light curves were corrected for \textit{K}-correction and Milky Way extinction using \snoopy, then interpolated with PISCOLA to achieve continuous rest-frame coverage.

Decomposing the NIR light curves, each one independently and using three PCA components ($C_0$, $C_1$, and $C_2$) explains $\sim90\%$ of the variability, with the first component explaining around $50\%$ of it. The $C_0$ component of the \J and \H bands ($C_1$ for the \Y band) shows that brighter secondary peaks also occur at later epochs and vice versa. With a somewhat opposing effect, the $C_1$ component of the \H band ($C_0$ for the \Y band) shows that brighter secondary peaks occur at earlier epochs, while this component mainly just contributes to the brightness for the \J band. A possible explanation for the NIR bands exhibiting different dominant behaviors is that the PCA components could be tracing different physical properties, such as the effect of \mni, metallicity, or \mfe on the time of recombination of iron-group elements and, subsequently, on the secondary peak. When comparing to observations and theoretical models, a difference in the elements abundance is seen for the different wavelengths ranges each of the bands probes, possibly explaining some of the differences seen with the PCA components.

When comparing the PCA coefficients ($p_0$ and $p_1$) against different light-curve parameters, some interesting trends are found. The NIR peak absolute magnitude correlates with $p_0$ for the \Y band, which could imply some correlation between secondary peak brightness and \mni. However, the same correlations are not seen for the \J and \H bands. Nonetheless, the optical peak brightness, which can be used as a proxy for \mni, correlates with $p_1$ for the \Y band and $p_0$ for the \J band, suggesting a possible diminishing influence of \mni in the secondary peak toward redder wavelengths.

We also compared the PCA coefficients against host-galaxy properties, in particular, stellar mass and color excess, but we found no clear correlations. However, we cannot say whether the environment does not affect the secondary NIR peak. A larger sample and, in particular, a comparison with other host-galaxy properties, such as the star formation rate and metallicity, would be needed to draw a strong conclusion in this regard.

The PCA decomposition was also applied on the NIR color curves of the \sneia, where over $50\%$ of the variance is explained by the first component ($C_0$). Although the physical interpretation of these is more complex, we still compared them against light-curve properties. We found correlations between the $C_0$ component and peak optical brightness (i.e., \mni) for all the NIR colors, showing that SNe with larger amounts of \mni have redder NIR colors. When comparing to the same host properties as before, we found a tentative correlation with \hostmass, where \sneia with redder NIR colors are found on less massive galaxies (potentially more metal-poor) environments. The comparison against \ebvhost shows some correlations, but mainly driven by two SNe and possibly due to the relatively low statistics.

Given the correlation found between $M^Y$ and $p_0$, we tested the standardization of the \Y-band light curves, finding a reduction in $\sigma_{\rm int}$ of $0.04$\,mag when correcting the \Y-band luminosity. Future research will allow us to investigate whether applying corrections at these wavelengths offers a genuine advantage, given the need for good photometric coverage, compared to using a single NIR epoch \citep{Stanishev2018, Müller-Bravo2022}, which benefits from increasing the statistical sample size of \sneia. The Vera C. Rubin Observatory Legacy Survey of Space and Time \citep{Ivezic2019} will provide \textit{y}-band photometry for SNe, which can be used as a test set.

NIR observations of \sneia provide a unique opportunity to study the physics of these transients and for doing cosmology. Although the number of objects observed in the NIR has been steadily increasing in the last decade, there is still a very limited number of these with high-cadence observations covering both NIR peaks. Surveys such as CSP present unique datasets to study these transients in detail given the rich photometric and spectroscopic coverage. Photometry from new low- and high-redshift surveys, such as the Aarhus-Barcelona FLOWS project\footnote{\url{https://flows.phys.au.dk/}}, Hawai'i Supernova Flows \citep{Do2025}, and JADES \citep{DeCoursey2025}, will further increase the numbers. This will be essential for making the most of future observations from Euclid and the Roman Space Telescope.

Finally, we would like to encourage the community to produce new and more sophisticated theoretical models of \sneia extending to NIR wavelengths. These models can be used to make comparisons against the increasing amount of data, helping constrain their physics and improve their usefulness as cosmological distance indicators.


\section*{Data Availability}

Most data used in this work form part of published articles. Published data can be shared upon request to the authors. Private data from \cspii will be published in an upcoming article (Suntzeff et al. in prep.).


\begin{acknowledgements}
The authors thank the anonymous referee for a constructive review that has improved the quality and content of this publication.
T.E.M.B. would like to thank Kate Maguire for very insightful discussions about the physical interpretation of the PCA components for the analysis.
T.E.M.B. and L.G. acknowledge financial support from the Spanish Ministerio de Ciencia e Innovaci\'on (MCIN), the Agencia Estatal de Investigaci\'on (AEI) 10.13039/501100011033, the European Social Fund (ESF) "Investing in your future, and the European Union Next Generation EU/PRTR funds under PID2020-115253GA-I00 HOSTFLOWS and PID2023-151307NB-I00 SNNEXT projects, the 2021 Juan de la Cierva program FJC2021-047124-I, and from Centro Superior de Investigaciones Cient\'ificas (CSIC) under PIE 20215AT016, ILINK23001, COOPB2304 projects, and the program Unidad de Excelencia Mar\'ia de Maeztu CEX2020-001058-M. 
T.E.M.B. is funded by Horizon Europe ERC grant no. 101125877.
M.D.S. and the FLOWS project is funded by the Independent Research Fund Denmark (IRFD, grant number  10.46540/2032-00022B) and by the Aarhus University Research Foundation (Nova grant 44369).
The work of the Carnegie Supernova Project has been supported by the NSF under the grants AST0306969, AST0607438, AST1008343, AST1613426, AST1613472 and AST613455.

\textit{Software}: 
\texttt{astropy} \citep{astropy, astropy2},
\texttt{corner} \citep{corner}, 
\texttt{emcee} \citep{emcee}, 
\texttt{extinction} \citep{Barbary2016-ext},
\texttt{matplotlib} \citep{matplotlib}, 
\texttt{numpy} \citep{numpy},
\texttt{pandas} \citep{pandas},
\texttt{peakutils} \citep{peakutils},
\texttt{PISCOLA} \citep{piscola},
\texttt{scikit-learn} \citep{scikit-learn}, 
\texttt{scipy} \citep{scipy},
\texttt{sfdmap}\footnote{\url{https://github.com/kbarbary/sfdmap}}, 
\texttt{SNooPy} \citep{Burns2011},
\texttt{tinygp} \citep{tinygp}
\end{acknowledgements}

\bibliographystyle{aa}
\bibliography{bibliography}

 \begin{appendix}

\section{Peculiar Type Ia supernovae}
\label{app:pec_SNeIa}

\begin{figure}[h]
    \includegraphics[width=\columnwidth]{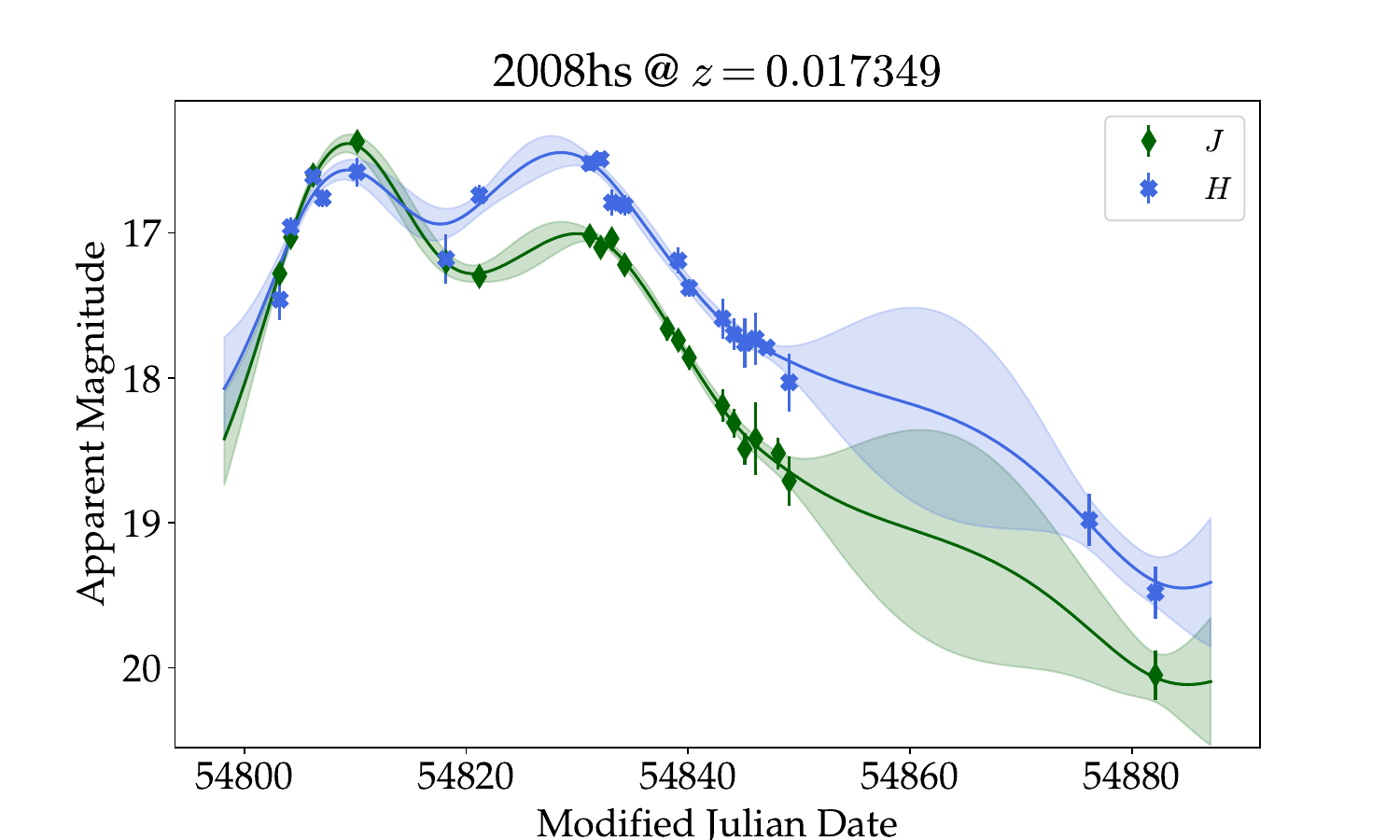}
    \includegraphics[width=\columnwidth]{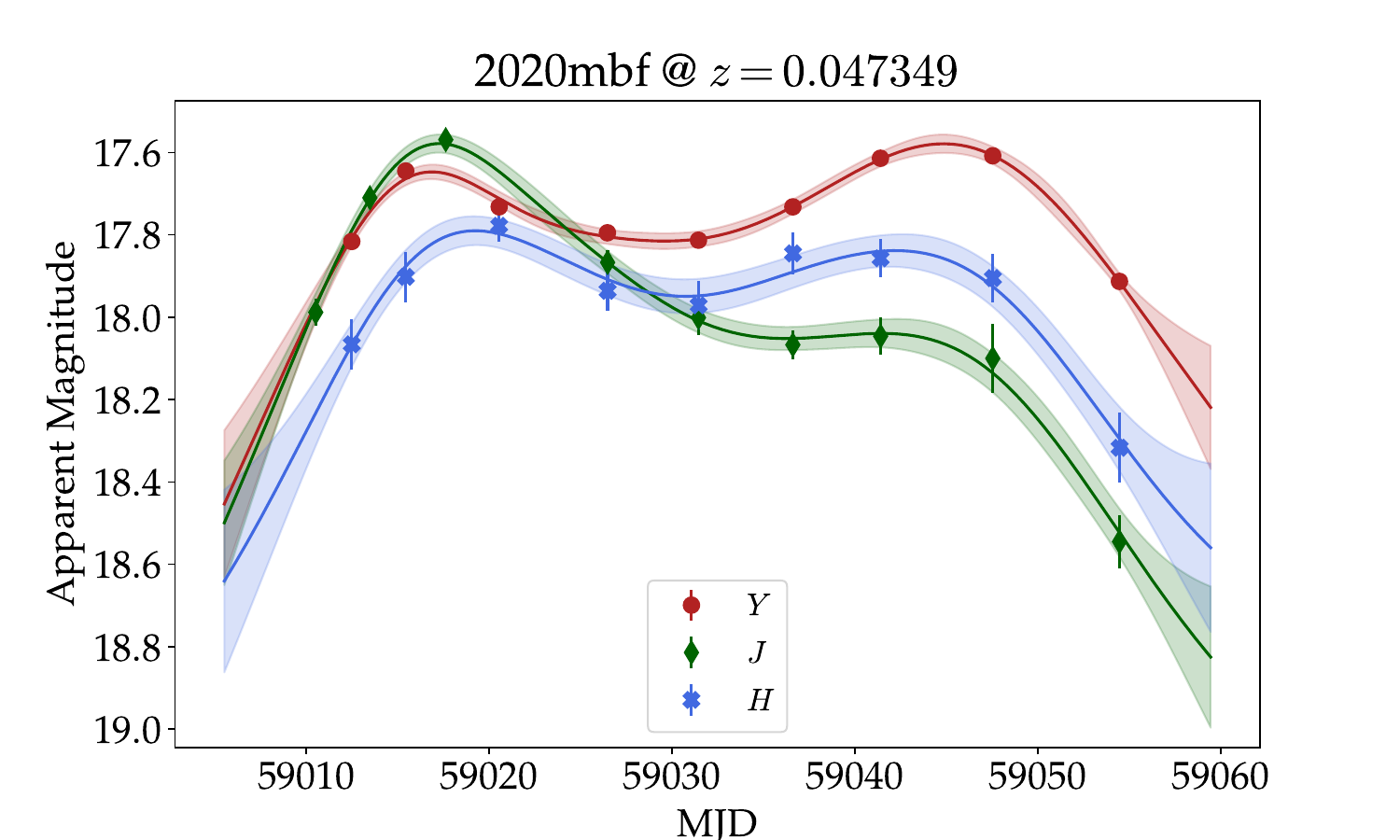}
    \caption{PISCOLA light-curve fits to SNe 2008hs and 2020mbf.}
    \label{fig:pec_SNeIa}
\end{figure}

The PISCOLA light-curve fits of SNe 2008hs (CfAIR2 sample) and 2020mbf (DEHVILS sample) are shown in Fig.~\ref{fig:pec_SNeIa}. SN 2008hs was originally classified as a normal \snia. However, given its relatively low \sbv value ($0.623$) and much earlier secondary NIR peak compared to the sample used in this analysis, this SN could be classified as a \lq transitional\rq\, \snia, such as iPTF13ebh \citep{Hsiao2015}. In the case of SN 2020mbf, it is very clear that the local NIR minimum is very shallow in all three bands. It is also of particular interest the \J-band light curve, as the difference in brightness between both peaks ($\sim0.6$\,mag) is much smaller compared to all other \sneia in this analysis ($>1.0$\,mag: see Fig.~\ref{fig:light_curves}). This flattening of the light curves is what defines 06bt-like SNe \citep{Foley2010}, although normally seen in the \textit{i} band. However, SN 2006mbf does not have coverage in this band.
A more extended analysis of these objects is beyond the scope of this work and we encourage their analysis by the community.

\section{NIR light-curve standardization with color components}
\label{app:color_cosmo}

\begin{figure}
    \includegraphics[width=\columnwidth]{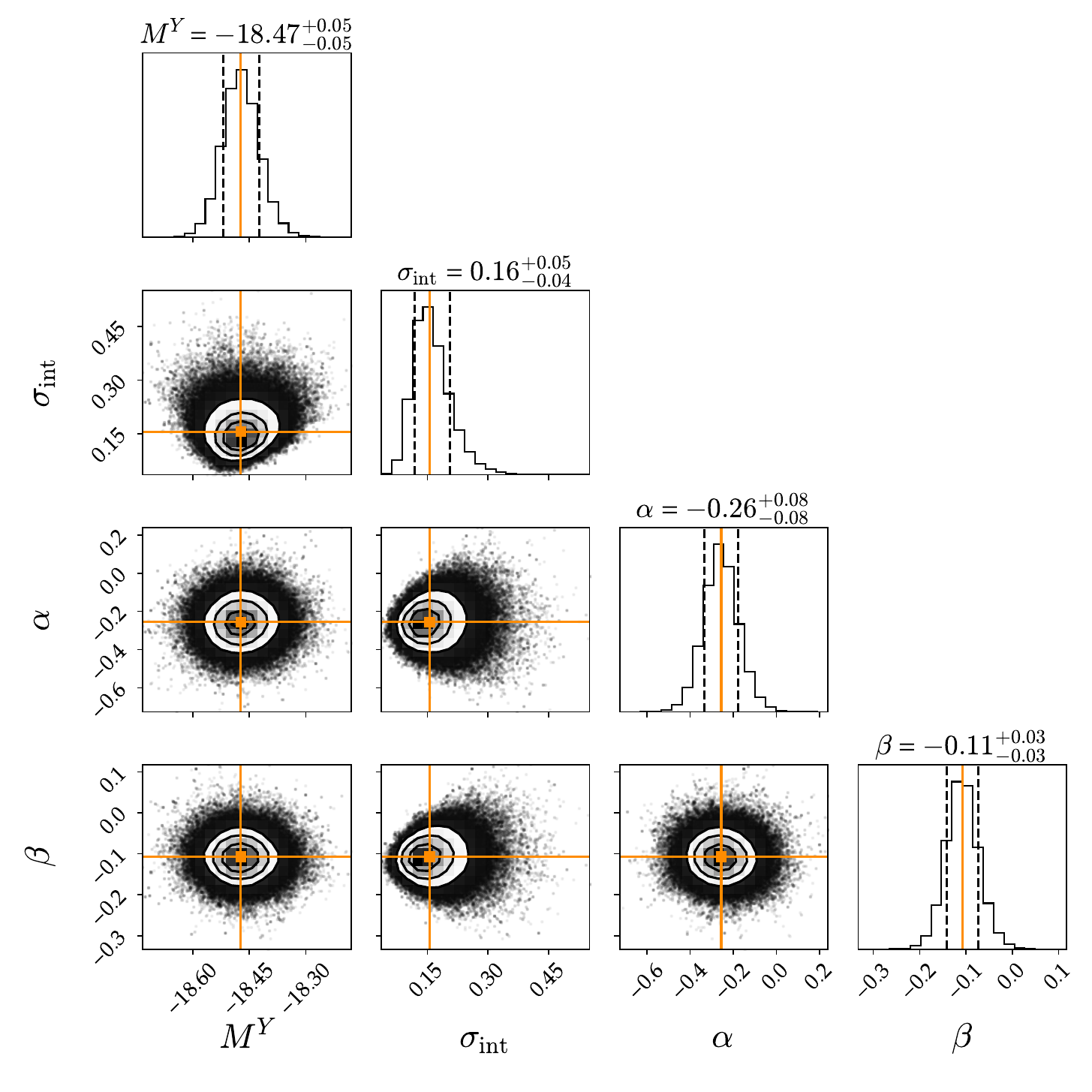}
    \caption{Same as Fig.~\ref{fig:Y_contours}, but using the PCA decomposition on the $(Y-H)$ color curves.}
    \label{fig:YH_contours}
\end{figure}

\begin{figure}
    \includegraphics[width=\columnwidth]{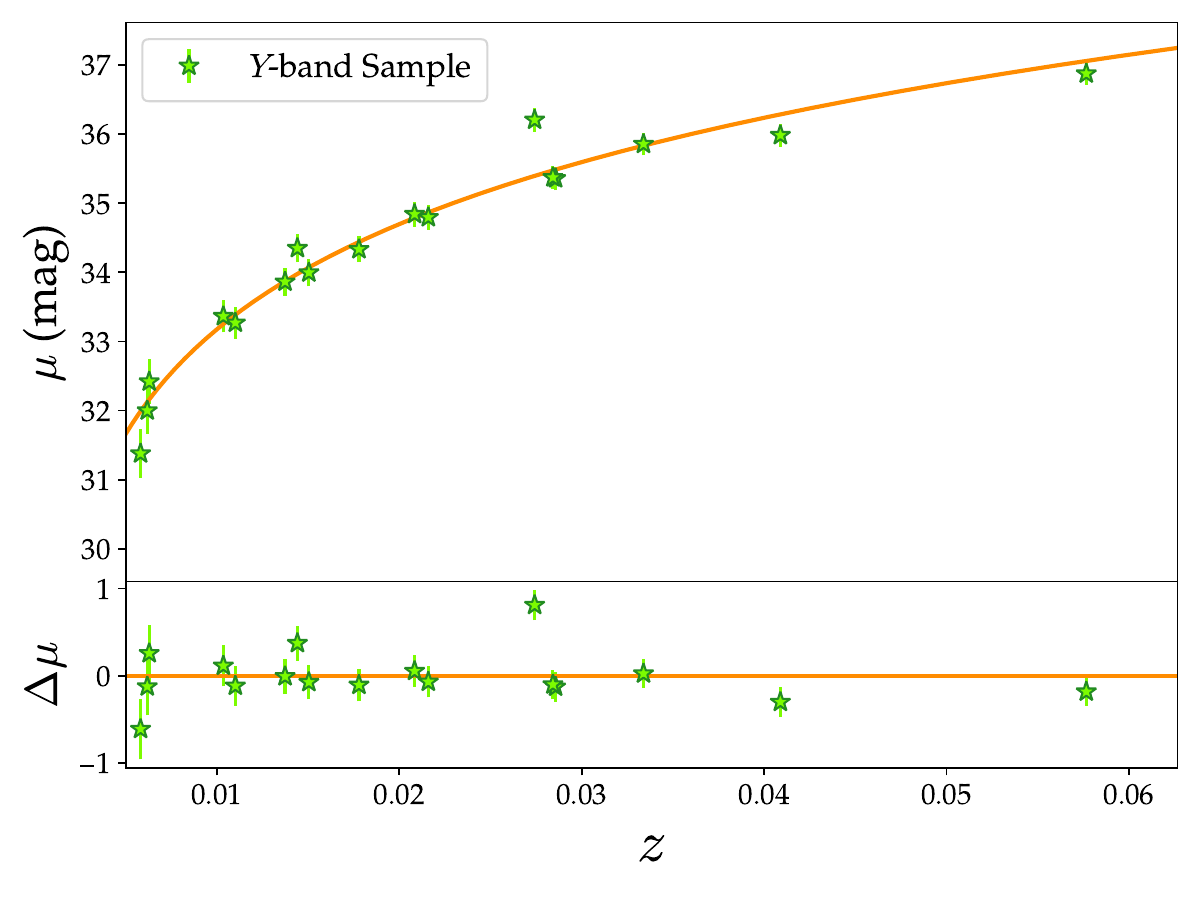}
    \caption{\Y-band Hubble diagram with the standardization from Fig.~\ref{fig:YH_contours}. The Hubble residual presents a RMS of $0.15$\,mag and a normalized median absolute deviation of $0.16$\,mag.}
    \label{fig:YH_HD}
\end{figure}

Figures~\ref{fig:YH_contours} and \ref{fig:YH_HD} show the corner plot and Hubble diagram, respectively, using the $(Y-H)$ color-curve PCA coefficients for the standardization (Sect.~\ref{subsec:color_decomposition}). In this case, the standardization is favored by around $3.3\sigma$ and $3.7\sigma$ significance for $p_0$ and $p_1$, respectively, while $\sigma_{\rm int}$ is significantly reduced from $0.27$\,mag to $0.16$\,mag. The reader should have in mind that this sample is smaller than the \Y-band only sample given the need of two bands, which explains the difference in $\sigma_{\rm int}$.

\end{appendix}
\end{document}